\documentclass[amsmath,amssymb,prb,twocolumn,superscriptaddress]{revtex4}

\usepackage[utf8]{inputenc}
\usepackage{amsmath}
\usepackage{graphicx}
\usepackage{hyperref}
\usepackage{enumerate}

\newcommand{\dipc}{Donostia International Physics Center (DIPC), E-20018 San Sebasti\'an, Spain}
\newcommand{\ikerbasque}{IKERBASQUE, Basque Foundation for Science, E-48013, Bilbao, Spain}

\newcommand{\Eqref}[1]{Eq.~(\ref{#1})}
\newcommand{\Figref}[1]{Fig.~\ref{#1}}

\begin{document}
\title{Magnetic frustration and fractionalization in oligo(indenoindenes)}

\author{Ricardo Ortiz}
\affiliation{\dipc}
\author{Geza Giedke}
\affiliation{\dipc}
\affiliation{\ikerbasque}
\author{Thomas Frederiksen}
\affiliation{\dipc}
\affiliation{\ikerbasque}
%\author{We-can-discuss Sequence}

\date{\today}

\begin{abstract}
Poly(indenoindenes) are $\pi$-conjugated ladder carbon polymers with alternating hexagons and pentagons hosting one unpaired electron for each five-membered ring in the open-shell limit. Here we study the main magnetic interactions that are present in finite oligo(indenoindenes) (OInIn), classifying the six possible isomers in two different classes of three isomers each. One class can be rationalized by frustrated $S=1/2$ Heisenberg chains, with ferromagnetic interactions between neighbour sites and antiferromagnetic interactions between the next neighbours. The other class is characterized by the more trivial antiferromagnetic order.
Employing several levels of theory we further show that the ground state of one of the isomers is a valence-bond solid (VBS) of ferromagnetic dimers ($S=1$). This is topologically similar to that of the Affleck-Kennedy-Lieb-Tasaki (AKLT) model, which is known to show fractional $S=1/2$ states at the edges.
\end{abstract}
\maketitle

The study of graphene-related systems as playground for realizing exotic phases of matter is a topic of intense research in modern physics. The observation of superconductivity in magic-angle twisted bilayer graphene\cite{cao2018unconventional}, or the discovery of fractional edge states in triangulene chains\cite{mishra2021observation}, would be fair examples of such claim.
Thus, as a generalization of the latter, open-shell nanographenes that host localized electrons can be used to design more complex architectures that mimic model and spin Hamiltonians, displaying the non-trivial physics of these models\cite{mishra2021observation,mishra2020,crommiescience,massivefermions2020,sun2020coupled,JPCLRuffieux2021,NachoNanostar2021,groning2018engineering,PRMOrtiz2022,toyAKLT,2Dtriang}. 

Inconveniently, the high reactivity inherent to unpaired electrons has prevented the realization of these systems for many decades. Solely recent experiments employing on-surface synthesis in ultra-high vacuum conditions has shown effective for obtaining such pristine open-shell molecules\cite{wang2016,Ruffieux16,pavlivcek2017synthesis,tailoringbondMishra2018}. Further characterization, by means of scanning tunneling microscopy (STM), has also probed the existence of local moments by measuring Kondo peaks\cite{NachoNat,li2019,ShantaJACS20} or inelastic steps\cite{rhombenes,mishra19b,ORTIZ2020100595}. To this day, this is a well-established technology, so there is a whole plethora of platforms with localized electrons and $\pi$-magnetism\cite{D0CS01060J,OtDiFr.22.Carbonbasednanostructures}.

Electron localization in nanographenes is related very often with states pinned at (or close to) the Fermi energy\cite{LADO201556}, which are originated by several sources like sublattice imbalance\cite{JFR07} or non-trivial topology\cite{koshino2014topological,louiejunction}.
These zero modes can interact by different exchange mechanisms\cite{ortiz19,superexchangejacob}, leading to either ferromagnetism or antiferromagnetism. The total spin quantum number of the ground state at half-filling can be predicted by the Lieb's theorem\cite{Lieb89} ($S=|N_A-N_B|/2$). This theorem was originally formulated for bipartite lattices (with $N_{A,B}$ sites belonging to the sublattices $A$ and $B$) and the Hubbard model, and later found to be in agreement with numerical results beyond this method\cite{cusinato2018electronic}.

\begin{figure}[h!]
 \centering
    \includegraphics[width=0.47\textwidth]{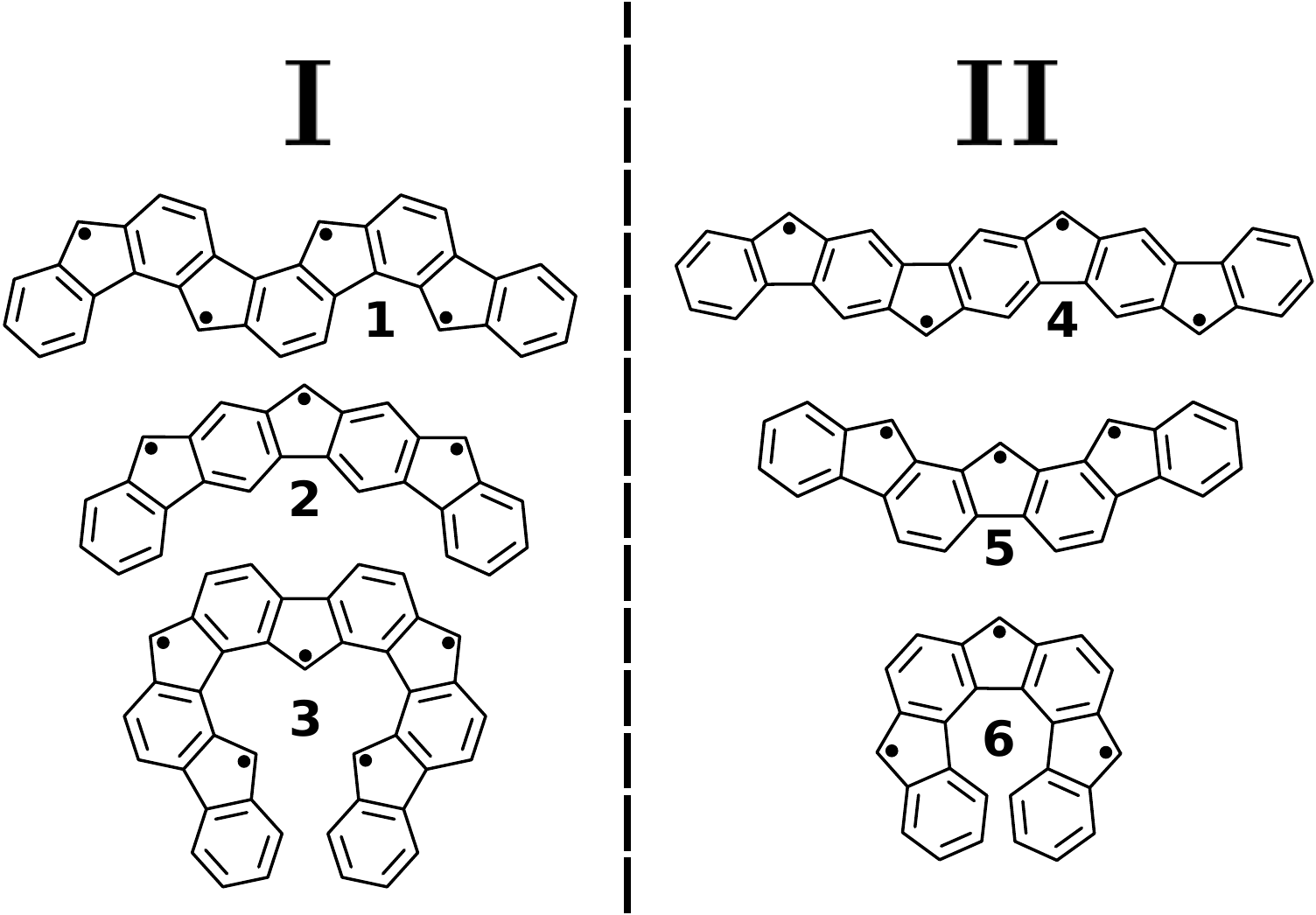}
\caption{Isomeric forms of the OInIn in the open-shell configuration. These systems would derive from the following molecules: isomers {\bf 1} and {\bf3} from indeno[1,2-a]fluorene, isomer {\bf 2} from indeno[2,1-b]fluorene, isomer {\bf 4} from indeno[1,2-b]fluorene, isomer {\bf 5} from indeno[2,1-a]fluorene and isomer {\bf 6} from indeno[1,2-c]fluorene. Any other isomeric system is just a combination of these. According to our analysis, the structures can be grouped into classes {\bf I} and {\bf II} according to different magnetic interactions.}
%\label{Structure}
\label{fig1}
\end{figure}

In the following we will primarily focus on non-bipartite systems, more specifically, finite conjugated ladder polymers that alternate hexagon and pentagon rings: the oligo(indenoindenes) (OInIn, \Figref{fig1}). The inclusion of $P$ pentagon rings induces frustration in the sublattices\cite{graphsatelites}, making Lieb's theorem no longer applicable, but one may expect exchange interactions to work similar as in bipartite lattices. For instance, two overlapping electrons present ferromagnetic (FM) exchange\cite{ortiz19,superexchangejacob} ($J^\mathrm{FM}\propto U$), whilst coupling through hopping leads to kinetic antiferromagnetic (AF) exchange\cite{andersonexchange,Ortiz18,superexchangejacob} ($J^\mathrm{AF}\propto  1/U$). It is worth to mention that there might be other interactions that can be important, \textit{i.e.}, Coulomb-driven exchange\cite{ortiz19,superexchangejacob}, but the former two will play a dominant role in the systems discussed here.

In \Figref{fig1}, we show the six possible isomers that can be drawn by keeping invariant the angle of two vectors that point in the vertex direction of adjacent pentagons.
Despite their high instability, experiments on several of these systems have been reported to this day, like the antiaromatic $P=2$ island\cite{transdimerexp} and oligomers\cite{transchainexp} of isomer {\bf 4}. In that case, it is expected by theory\cite{JACSN2,C4CP03169E}, and confirmed by atomic force microscopy (AFM) experiments\cite{transdimerexp}, that the shortest island is closed-shell, but the ground state develops magnetism when the system is enlarged according to calculations\cite{transchainexp,yamane2018open,fukuda2017theoretical}.
The synthesis of a $C_3$ symmetric derivative (truxene-5,10,15-triyl) with $S=3/2$ ground state is also confirmed\cite{truxene,truxenewang}, as well as polymers of $P=2$ islands\cite{chainisomer5} of isomer {\bf 2}, and polymers of $P=2$ islands of isomer {\bf 4} with intercalations\cite{intercalationsJACS} of isomer {\bf 5}.

Drawing the maximum number of Clar sextets in OInIn leaves one unpaired electron per pentagon\cite{hu2016toward}, justifying their radical character. According to this, we argue that such indenofluorene derivatives, with $P$ pentagons and $P+1$ hexagons, may be understood as effective electron chains of size $P$. This picture corresponds to the open-shell limit, and it is convenient for our purpose in this article, although a more realistic description would consist of a combination of different configurations that reduces the radical character\cite{yeh2016role}. This is of especial relevance for some $P=2$ islands that are closed-shell\cite{JACSN2,C4CP03169E}, but simplifying the system to electron chains allows us to study the exchange mechanisms at a fundamental level, which also explains the physics of the shortest molecules.

\begin{figure*}
 \centering
    \includegraphics[width=1\textwidth]{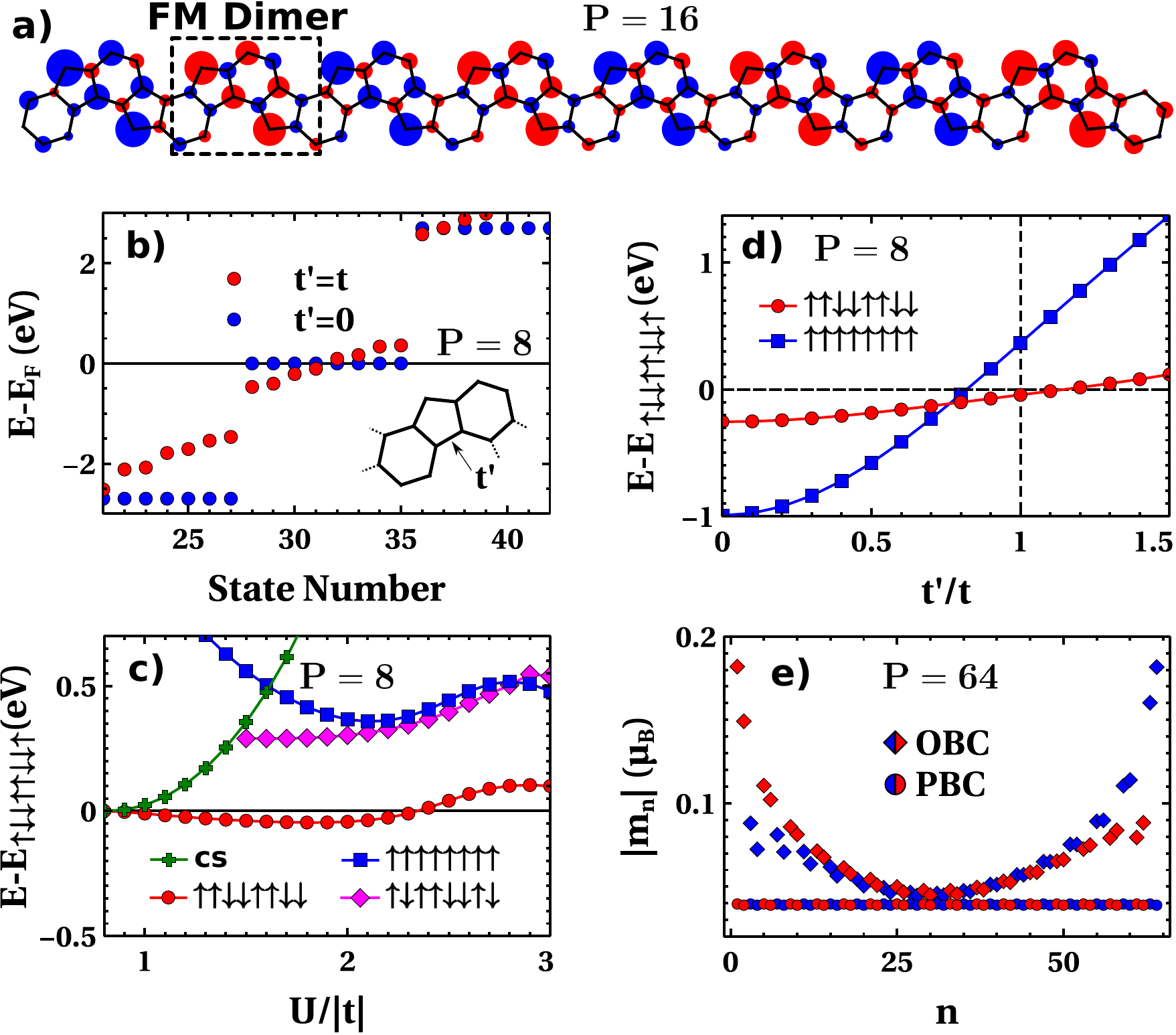}
\caption{Magnetic results obtained with MF-Hubbard model for isomer {\bf 1}. a) Magnetic moments $m_i$ per carbon site $i$ of the lowest energy solution calculated for $U=1.5|t|$, isomer {\bf 1} with $P=16$ pentagons. Color stands for the sign and the area for the relative magnitude of the moment. Double bonds and H atoms are removed for clarity. b) Single-particle spectra around the Fermi energy with $t'=0$ and $t'=t$ for isomer {\bf 1} with $P=8$, calculated with the tight-binding model ($t=-2.7$ eV). 
c) Energy difference for different magnetic solutions (labeled as $\sigma_1\sigma_2...\sigma_P$, where $\sigma_n$ is the sign of the moment at the $n^\mathrm{th}$ pentagon vertex) and a solution with, apparently, FM dimers throughout the molecule except the edges ($\uparrow\downarrow\downarrow\uparrow\uparrow\downarrow\downarrow\uparrow$, hereafter broken dimers), converged with MF-Hubbard as a function of $U$ for isomer {\bf 1} with $P=8$. CS stands for closed-shell. 
d) Energy difference for the $\uparrow\uparrow\downarrow\downarrow\uparrow\uparrow\downarrow\downarrow$ (FM dimers) and $\uparrow\uparrow\uparrow\uparrow\uparrow\uparrow\uparrow\uparrow$ (FM) solutions and the broken dimers as a function of $t'$ for isomer {\bf 1} with $P=8$, calculated with MF-Hubbard model and $U=2|t|$.
e) Plots of the absolute values of the magnetic moments at each pentagon vertex atom $n$ of the lowest energy solution (FM dimers) for isomer {\bf 1} with $U=0.85|t|$ and $P=64$, considering both periodic (PBC) and open (OBC) boundary conditions. The red/blue colors stand for the sign of each moment. The formation of local magnetic edge moments in the OBC case is clearly seen.}
\label{fig2}
\end{figure*}

Thus, in this work two things are done. First, we consider that the OInIn can be interpreted, in the tight-binding approximation, as bipartite systems plus a hopping $t'$ that closes the pentagons \cite{kertesz1995structure}. Then, when interactions are considered, 
we show that isomer {\bf 1} presents a ground state in the vicinity of $t'=t$ that consists in $P/2$ FM dimers with effective $S=1$ quantum number each (\Figref{fig2}a). Second, in order to explain these results, we perform an analysis of the different magnetic exchanges that may be present in the molecule. By this, we manage to classify the six isomers in two different groups: antiferromagnets (isomers {\bf 4}, {\bf 5} and {\bf 6}) and frustrated chains with antiferromagnetic second neighbours interactions (isomers {\bf 1}, {\bf 2} and {\bf 3}).

\textit{Methods.} 
In order to obtain our results, we consider electron interactions by means of the Hubbard model, that we solve using two approaches: a collinear mean-field (MF) approximation and an exact diagonalization with a complete active space CAS($N_e,N_o$), where $N_e$ ($N_o$) refers to the number of electrons (single-particle orbitals) included. Then, we compare the results of isomer {\bf 1} with those from density functional theory (DFT), where we employed the Quantum Espresso\cite{giannozzi2009quantum,giannozzi2017advanced,giannozzi2020} and ORCA\cite{orcageneric} packages with the PBE-GGA\cite{perdew1996generalized} and PBE0\cite{hybrids1,PBE0hybrid} exchange-correlation functionals, respectively (see supp.~mat.).

\textit{Results.} %In the following, we present the results obtained with the MF-Hubbard model for isomer {\bf 1}. 
The origin of magnetism in the OInIn lies in the unpaired electrons, that are mainly localized at the pentagons. These can be visualized in a simple non-interacting tight-binding approximation, where $P$ single-particle states are present inside a big gap in isomer {\bf 1} (\Figref{fig2}b). 
As we can see, such in-gap states can be understood as the hybridized zero modes of the bipartite molecule with $t'=0$ and sublattice imbalance $|N_A-N_B|=P$. Apparently, this hybridization is not enough to prevent magnetism, and local moments appear in a MF-Hubbard model calculation for a $P=8$ molecule when $U>0.8|t|$ (\Figref{fig2}c). 
Counter-intuitively, these pentagon moments do not display either ferromagnetism or antiferromagnetism, but instead they organize as FM dimers with AF order, distributed all over the chain (\Figref{fig2}a). The stability of these dimers can be tested by modulating $t'$, where they are the ground state in isomer {\bf 1} when $0.75< t'/t< 1.15$, showing that a certain range of hybridization is key for their existence (\Figref{fig2}d).

In \Figref{fig2}c, we show the energy difference between some magnetic solutions of an isomer {\bf 1} $P=8$ molecule and the $\uparrow\downarrow\downarrow\uparrow\uparrow\downarrow\downarrow\uparrow$ configuration as a function of the on-site Coulomb repulsion $U$. After the system turns magnetic, the dimers dominate with a maximum energy difference of $46$ meV for $U\approx1.8|t|$. 
This representation reveals the stabilization inherent to the formation of the dimers, since a solution $\uparrow\downarrow\uparrow\uparrow\downarrow\downarrow\uparrow\downarrow$ with just two dimers is higher in energy and $\uparrow\downarrow\uparrow\downarrow\uparrow\downarrow\uparrow\downarrow$ could not be converged.
If we keep increasing $U$, the dimers stability is compromised, so they are the ground state for  $0.8|t|< U<2.3|t|$, which lies in the range that is usually taken to be physically relevant for nanographenes\cite{Yazyev_2010}. 
Additionally, we compute the FM phase with $S=P/2$, which is a higher energy solution until $U\approx 3.5|t|$ (not shown) that becomes the ground state.

In addition, we also performed DFT calculations for isomer {\bf 1} with different lengths and density functionals (see Table~\ref{table1}). 
In all cases, the FM dimers were the most stable solution. Nonetheless, with the PBE-GGA functional, for the $P=4$ molecule the difference with $\uparrow\downarrow\downarrow\uparrow$ is just of $2.2$ meV, and $2.1$ meV with the closed-shell solution, which may shed doubts about a spin-polarized configuration as the lowest energy state. However, for either larger polymers, or with the hybrid PBE0 functional, this energy separation increased, which validates the FM dimers as the ground state, as we obtained with the MF-Hubbard model. 

\begin{figure}
 \centering
    \includegraphics[width=0.47\textwidth]{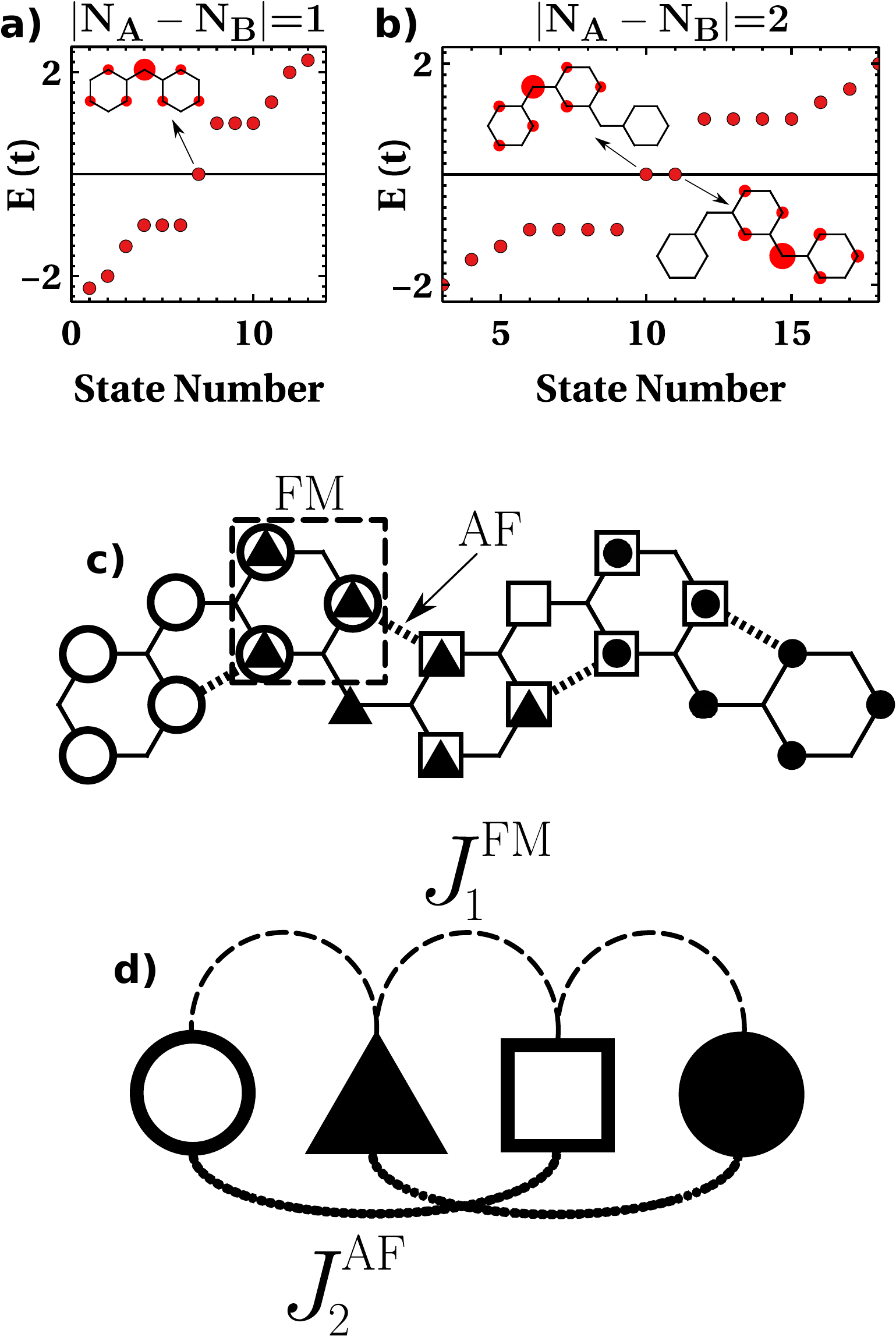}
\caption{Analysis of the different magnetic exchanges present in OInIn. a) and b) are the non-interacting tight-binding spectra of two OInIn with $P=1$ and $P=2$, both with $t'=0$. The absence of pentagons allows the bipartite classification of the lattice, and the sublattice imbalance of each molecule entails the presence of zero modes. The insets at each panel stand for the probability distribution of these zero modes. c) Sketch of the zero modes distribution of $t'=0$ molecules on a lattice of isomer {\bf 1}. There are two ways these zero modes couple considering just first-neighbours hopping, which leads to a FM first-neighbours interaction and first/second-neighbours AF exchange when $t'\neq 0$. d) Effective spin-1/2 model that the isomers from class {\bf I} can be mapped to according to this analysis.}
\label{fig3}
\end{figure}

We want to point out one further remarkable feature of these polymers. In \Figref{fig2}e we represent the absolute value of the magnetic moment at each pentagon vertex $n$ for a large isomer {\bf 1} molecule with $P=64$, calculated with the MF-Hubbard model for $U=0.85|t|$. With periodic boundary conditions (PBC), positive and negative moments of equal magnitude alternate every two pentagons. 
When the edges are present, we can see that, whilst they keep the same dimeric pattern, the magnetic moments get strongly localized at the termini, suggesting non-trivial topology.

In order to understand the physics behind this magnetic behaviour, it is useful to consider these systems as just bipartite lattices composed of hexagons, linked by an extra carbon atom, plus an additional hopping that completes the ladder backbone\cite{kertesz1995structure}. Then, for $P=1$ and $t'=0$, we have a nanographene that consists of two hexagons with one carbon that serves as linker. Such molecule displays sublattice imbalance and one zero mode mainly localized at the carbon atom that links the hexagons (\Figref{fig3}a). 
All the isomers are formed depending on the position and the rotation of the bond that connects to the next linker.
For instance, the molecule at the inset of \Figref{fig3}b corresponds to isomer {\bf 1} with $P=2$ and $t'=0$. In this case, the second linker adds an extra atom to the majority sublattice, and therefore $|N_A - N_B| = 2$. Interestingly, the wave functions of the degenerate zero modes can be chosen to be localized in a similar way as that of the previous $P=1$ molecule.

As a consequence, an isomer {\bf 1} polymer with $P$ pentagons and $t'=0$ presents $P$ zero modes, localized each one of them at one linker and, to a lesser degree, at the adjacent hexagons. This is schematically represented in \Figref{fig3}c, from which we can infer three main magnetic interactions when $t'$ is included in the picture. First, two first-neighbouring unpaired electrons share one hexagon, where their wave functions overlap, leading to a FM Hund's interaction\cite{superexchangejacob} that scales with $U$. Second, $t'$ hopping promotes a kinetic Anderson AF exchange\cite{superexchangejacob}, proportional to $1/U$, that happens between first- and second-neighbours. Summarizing: isomer {\bf 1} presents competing AF and FM first-neighbours interactions and AF second-neighbours interaction between the unpaired electrons. 

 \begin{table}[htp]
 \begin{center}
    \begin{tabular}{| l | c | c | c | c | c |}
    \hline
 \multicolumn{5}{|c|}{PBE-GGA} \\
 \hline
    $\Delta E$ (meV) & $P = 4$ & $P = 6$ &  $P = 8$ & 1D \\ \hline\hline
    Broken Dimers & 2.2 & 38.5 & 18.4 & -\\
    Closed-Shell & 2.1 & 118.6  & 41.1 & 28.6 \\
    FM ($S=P/2$) & 654.4 & 854.9  & 1316.5 & 741.8 \\
    \hline\hline
    
    \multicolumn{5}{|c|}{PBE0} \\
 \hline
    $\Delta E$ (meV) & $P = 4$ & $P = 6$ &  $P = 8$ & 1D \\ \hline\hline
    Broken Dimers & 93.6 & 118.2 & 133.9 & -\\
    Closed-Shell & 504.7 & 1153.2 & 1084.5 & -\\
    FM ($S=P/2$) & 293.1 & 511.1  & 741.1 & - \\
    \hline
    \end{tabular}
    \label{table1}
\end{center}
\caption{Energy differences $\Delta E$ for the different magnetic solutions of isomer {\bf 1}, relative to that of the FM dimers, calculated with DFT for three molecular lengths and two different density functionals (PBE-GGA and PBE0), along with the periodic 1D polymer with the PBE-GGA functional. The relaxed unit cell of the 1D polymer included four pentagons (see sup. mat.), and the energies for this system are referred to that unit cell. The relaxed geometry of the FM dimers solution was used for calculating all the magnetic phases.}\label{table1}
 \end{table} 
 
\begin{figure}
 \centering
    \includegraphics[width=0.48\textwidth]{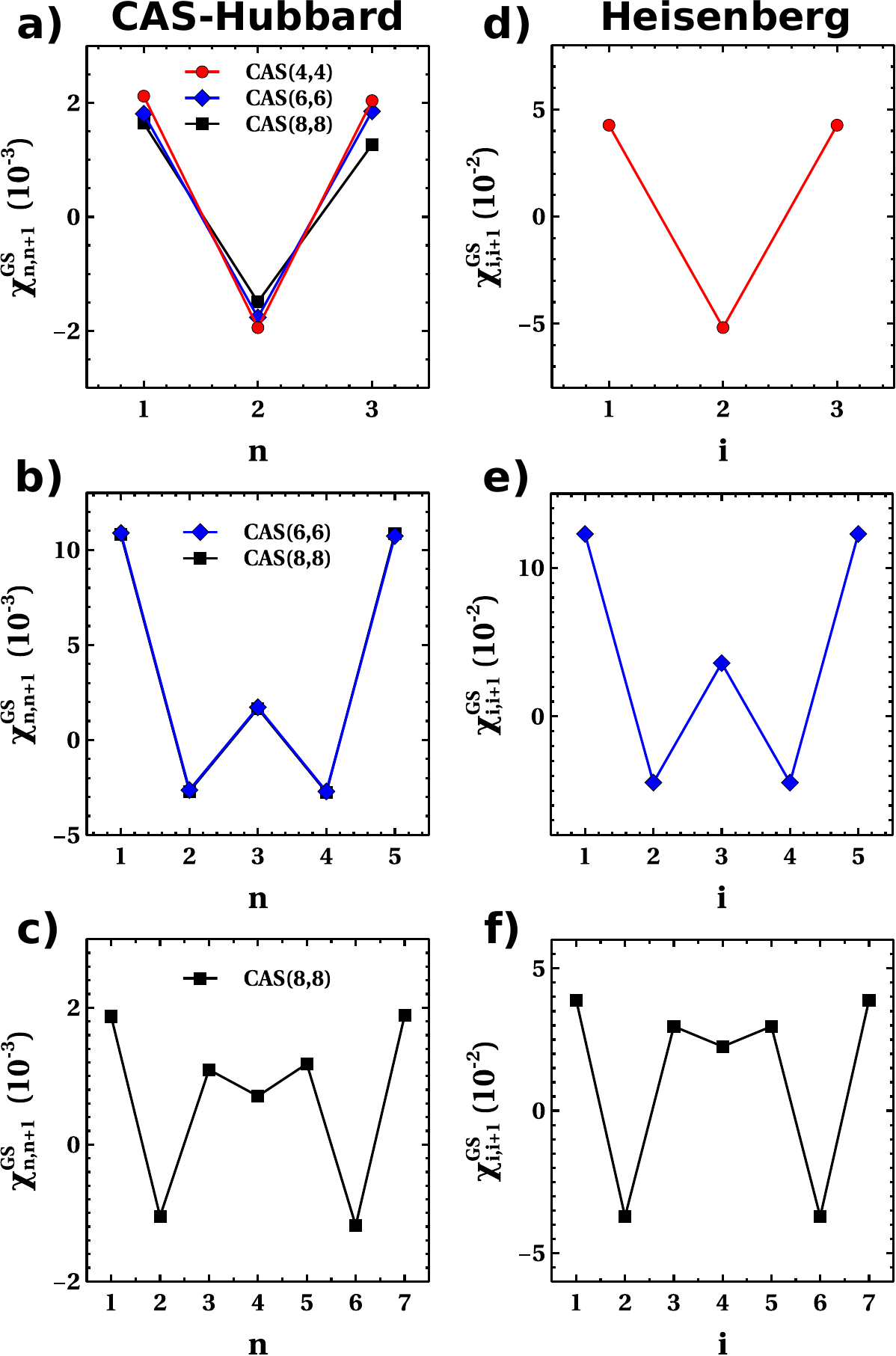}
\caption{Spin-correlator analysis for isomer {\bf 1} molecules, comparing with the frustrated FM Heisenberg chain with OBC.
a), b) and c) are the spin correlators between the vertex atom of first-neighbouring pentagons for molecules with different sizes and $U=1.5|t|$. d), e) and f) are the spin correlators between first-neighbouring $S=1/2$ spins of the frustrated FM chain with second-neighbours AF exchange and $J_{1}^\mathrm{FM} = -2J_{2}^\mathrm{AF}$.}
\label{fig4}
\end{figure}

The fact that we converge FM dimers as the lowest energy phase of isomer {\bf 1} is not trivial, since dimerization may happen as a result of magnetic frustration. Actually, two examples where dimerization occurs in the ground state are the frustrated $S=1/2$ Heisenberg chains with second-neighbours AF exchange and either AF or FM first-neighbours interactions\cite{majumdar1969next,majumdar1970antiferromagnetic,HaldaneMG,van1980exact,spindimers,furusakichain,agrapidis2019coexistence}: 
\begin{equation}
{\cal H} = J_1\sum_{\langle i,j\rangle} \vec{S}_i\cdot \vec{S}_j +J_2^\mathrm{AF}\sum_{\langle\!\langle i,j\rangle\!\rangle} \vec{S}_i\cdot \vec{S}_j,
\label{frustheise}
\end{equation}
where $\vec{S}_i$ is the spin operator vector and $\langle i,j\rangle$ ($\langle\!\langle i,j\rangle\!\rangle$) runs over first (second) neighbouring spins with $J_1$ ($J_2^\mathrm{AF}$) magnetic exchange.

Hence, \Eqref{frustheise} has two possibilities. First, when $J_1\equiv J_1^\mathrm{AF}>0$, the model can be exactly solved at the Majumdar-Ghosh point\cite{majumdar1969next,majumdar1970antiferromagnetic,van1980exact,HaldaneMG,spindimers} ($J_1^\mathrm{AF} = 2J_2^\mathrm{AF}$), and the ground state consists of a valence-bond solid (VBS) of singlets that served as inspiration of the AKLT model\cite{akltpaper}. 
Second, considering PBC, in the $J_1\equiv J_1^\mathrm{FM}<0$ case\cite{furusakichain,agrapidis2019coexistence} we find two  regimes: 
$|J_1^\mathrm{FM}/J_2^\mathrm{AF}|>4$, where the ground state is the FM solution, and $|J_1^\mathrm{FM}/J_2^\mathrm{AF}|<4$, where frustration promotes spontaneous symmetry breaking, through a process of order by disorder\cite{villain1980order}, that leads to a VBS of FM dimers\cite{furusakichain,agrapidis2019coexistence}. 
Interestingly, since these FM dimers effectively behave as $S=1$ spins, this phase has been theorized to have a hidden topological order similar to the AKLT model, showing a spin gap and edge-spin fractionalization\cite{agrapidis2019coexistence}.

Therefore, considering that when $U$ increases the FM eventually surpasses the AF first-neighbours interaction, isomer {\bf 1} can be described by a frustrated FM $S=1/2$ Heisenberg chain.
This scenario, illustrated in \Figref{fig3}d, along with
the results shown in \Figref{fig2}, lead us to suggest that the FM dimers shown so far are, indeed, such VBS, and the edge magnetization originates in the fractionalization of an effective Haldane chain\cite{akltpaper}.

In order to confirm this statement, we calculate spin correlators with a CAS-Hubbard model between the different unpaired electrons in isomer {\bf 1}, and compare them with those between neighbouring spins in the frustrated FM Heisenberg chain (\Figref{fig4}).
To do this, we first compute the multiconfigurational eigenstates from an exact Hubbard model of isomer {\bf 1} with $U=1.5|t|$. Since the VBS from polymers with $P=4m$, where $m$ is an integer, must be a singlet, we cannot use $\langle S_z(i)\rangle$ to assess the magnetic moment. Instead, we compute the spin correlator between each pentagon-vertex that are separated by one hexagon ($n$, $n+1$):
\begin{equation}
\chi^{\Psi}_{i,j} = \langle \Psi | S_z(i)S_z(j) | \Psi\rangle,
\end{equation}
where $S_z(i)$ is the $z$-component of spin operator at site $i$ and $\Psi$ is a many-body wave function.

In \Figref{fig4}a-c we show the correlators between the adjacent unpaired electrons in isomer {\bf 1} for different lengths. $\Psi$ is chosen to be the ground state, which changes spin multiplicity depending on $U$ (supp. mat.). Since we are looking for the FM dimers phase, the selected $U$ was inside a range where the polymers with $P=4m$ ($P=4m+2$) have an $S=0$ ($S=1$) ground state. These different spin quantum numbers are imposed according to whether the VBS presents even or odd number of dimers. Interestingly, for large $U$ the ground state has $S=P/2$, as it happens for the frustrated Heisenberg chain when $|J_1^\mathrm{FM}/J_2^\mathrm{AF}|>4$.

In the left panels of \Figref{fig4} we can clearly see the formation of the FM dimers, either by an alternation of signs in the spin correlators or by positive correlators with alternating strength \cite{furusakichain}, as in the inner pentagons of the $P=8$ molecule. These results are in qualitative agreement with the correlators between neighbouring spins of the frustrated FM Heisenberg chain (\Figref{fig4}d-f), which indicates that isomer {\bf 1} is correctly captured by this spin model.

\begin{figure}
 \centering
    \includegraphics[width=0.48\textwidth]{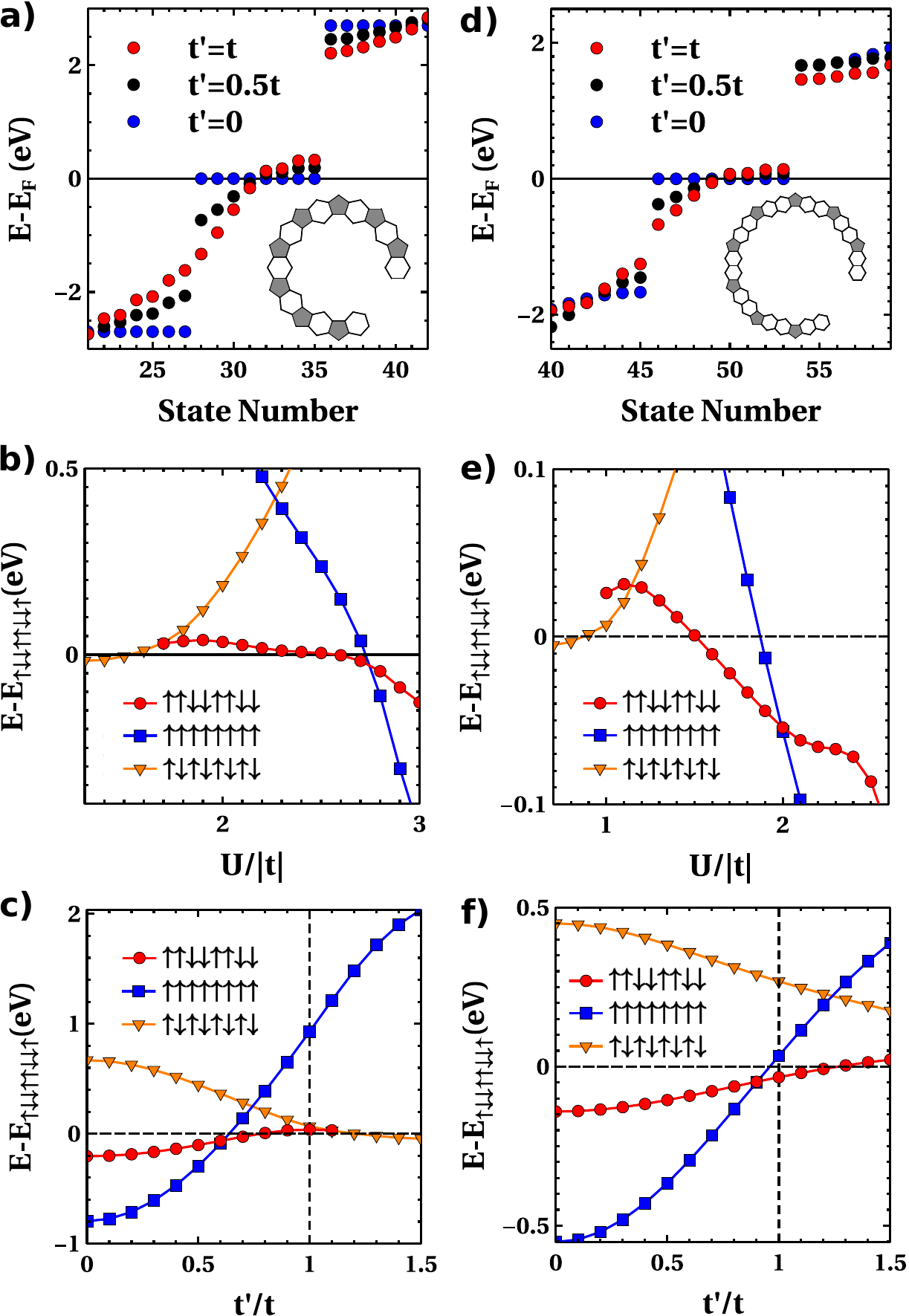}
\caption{Study of the effects of radical hybridization on isomer {\bf 2}. a), d) Non-interacting spectra for different $t'$, with $t=-2.7$ eV. b), e) Energy differences between different magnetic solutions as a function of $U$ calculated with MF-Hubbard. c), f) Energy differences between different magnetic solutions as a function of $t'$, calculated with the MF-Hubbard model and $U=1.8|t|$. Left and right panels are for a molecule of isomer {\bf 2} with $P=8$ with one and two hexagons separating the pentagons, respectively (insets of top panels). }
\label{fig5}
\end{figure}

We want to stress in the possibility of doing a similar analysis than that of \Figref{fig3} for the other isomers. In this sense, we may anticipate their magnetic properties by means of their sublattice imbalance when $t'=0$, that leads to the classification in \Figref{fig1}. 
Isomers {\bf 2} and {\bf 3} have sublattice imbalance, then the wave functions of neighbouring unpaired electrons overlap and belong to the same class as isomer {\bf 1}. On the other hand, isomers {\bf 4}, {\bf 5} and {\bf 6}, with $|N_A - N_B| =0$, do not present this feature, and unpaired electrons are connected by first-neighbours hopping. The closure of the pentagon does not add here any significant interaction, hence these isomers can not be described by a frustrated chain, but by a tight-binding chain instead (see supp. mat.). Thus, the energy separation between the frontier molecular orbitals decreases with the length, that stabilizes the AF spin-polarized solution, in agreement with the notion that $P=2$ molecules of class {\bf II} are closed-shell\cite{JACSN2,tobepure} that become magnetic when the size of the polymer is  increased\cite{transdimerexp,yamane2018open,fukuda2017theoretical}.

Class {\bf I} isomers, on the other hand, as we have shown, are more complicated. 
In the literature\cite{dressler2017synthesis}, the smallest island of isomer {\bf 1} is expected to be a triplet, which agrees with our results, but isomer {\bf 2} with $P=2$ is predicted to be an open-shell singlet\cite{shimizu2013indeno,tobepure}.
This can be explained if we consider that the larger hybridization in isomer {\bf 2} (\Figref{fig5}a) strengthens the AF first-neighbours interactions enough to not be surpassed by the FM exchange. If that is the situation, the system can be rationalized with an $S=1/2$ AF Heisenberg chain with AF second-neighbours exchange, that justifies the $S=0$ solution of the $P=2$ island, and anticipates different physics for isomer {\bf2} despite it belonging to the same class than isomer {\bf 1}.

This model, for larger chains and away of the Majumdar-Ghosh point, exhibits two distinct phases. When $J_1^\mathrm{AF} > 2J_2^\mathrm{AF}$, the numerical ground state consists in singlet dimers with a weak AF correlation, that is clearly seen in calculations of spin correlators calculated with the CAS-Hubbard model (supp. mat.). On the other hand, when $J_1^\mathrm{AF} < 2J_2^\mathrm{AF}$, the singlet dimers are ferromagnetically correlated\cite{white1996dimerization}, which looks like the broken dimers $\uparrow\downarrow\downarrow\uparrow\uparrow\downarrow\downarrow\uparrow$ ground state from MF-Hubbard of isomer {\bf 2} with $P=8$ at $1.53|t|< U<2.57|t|$.
However, the effective $\widetilde{U}/|\widetilde{t}|$ ratio\cite{Ortiz18}  for isomer {\bf 1} ({\bf 2}) is $16.5$ ($1.8$) (where $\widetilde{t}=\langle z_1|{\cal H}_t|z_2\rangle$, $\widetilde{U}=U\sum_i |z_{1,2}(i)|^4$, and $z_{1,2}$ are two adjacent zero modes of molecules with $t'=0$ and $U=1.5|t|$). Since the open-shell picture gets obscured\cite{superexchangejacob} when $\widetilde{U}\sim|\widetilde{t}|$, then the description by a spin chain for isomer {\bf 2} may be compromised, concluding that further theoretical effort is needed to actually assess its ground state.

Finally, we explore a little further the idea that the FM dimers are unstable due to electron hybridization. As we can see in \Figref{fig5}c, by decreasing $t'$ it is possible to obtain the FM dimers as the ground state, but this modulation of $t'$ seems unrealistic for a real carbon bond. 
Instead, by adding an additional hexagon between each pair of pentagons, the lower single-particle gap opens for $t'=t$ (\Figref{fig5}d) and the FM dimers are now the ground state for a broader window of $U$ and in the $t'\approx t$ vicinity (\Figref{fig5}e,f).
These results show that, by decreasing the hybridization, the description by the frustrated FM chain becomes more feasible. Then, the FM dimers get stabilized, but also the FM phase, which is the ground state for $U>2.0|t|$.

\textit{Conclusions.} We have presented the FM dimers as the ground state of isomer {\bf 1}, with characteristic magnetic localization at the termini that suggests fractionalization.
To understand this behaviour, we performed an analysis of the magnetic interactions that classifies the isomers into either class {\bf I} with competing FM and AF first-neighbours exchange and second-neighbours AF exchange, or class {\bf II} with first-neighbours AF exchange only.
The sign and range of these magnetic interactions are enough to explain the results obtained for isomer {\bf 1}, since the FM dimers can be identified as the ground state of the $S=1/2$ FM Heisenberg chain with AF second-neighbours exchange.

\textit{Acknowledgments.} We thank Pavel Jel{\'i}nek for fruitful discussions. This work was funded by the Spanish MCIN/AEI/ 10.13039/501100011033 (PID2020-115406GB-I00), the Basque Department of Education (IKUR Strategy under the collaboration agreement between Ikerbasque and DIPC), and the European Union’s Horizon 2020 (FET-Open project SPRING Grant No.~863098).

\bibliographystyle{apsrev-title}
\bibliography{references.bib}

\begin{thebibliography}{69}
\expandafter\ifx\csname natexlab\endcsname\relax\def\natexlab#1{#1}\fi
\expandafter\ifx\csname bibnamefont\endcsname\relax
  \def\bibnamefont#1{#1}\fi
\expandafter\ifx\csname bibfnamefont\endcsname\relax
  \def\bibfnamefont#1{#1}\fi
\expandafter\ifx\csname citenamefont\endcsname\relax
  \def\citenamefont#1{#1}\fi
\expandafter\ifx\csname url\endcsname\relax
  \def\url#1{\texttt{#1}}\fi
\expandafter\ifx\csname urlprefix\endcsname\relax\def\urlprefix{URL }\fi
\providecommand{\bibinfo}[2]{#2}
\providecommand{\eprint}[2][]{\url{#2}}

\bibitem[{\citenamefont{Cao et~al.}(2018)\citenamefont{Cao, Fatemi, Fang,
  Watanabe, Taniguchi, Kaxiras, and Jarillo-Herrero}}]{cao2018unconventional}
\bibinfo{author}{\bibfnamefont{Y.}~\bibnamefont{Cao}},
  \bibinfo{author}{\bibfnamefont{V.}~\bibnamefont{Fatemi}},
  \bibinfo{author}{\bibfnamefont{S.}~\bibnamefont{Fang}},
  \bibinfo{author}{\bibfnamefont{K.}~\bibnamefont{Watanabe}},
  \bibinfo{author}{\bibfnamefont{T.}~\bibnamefont{Taniguchi}},
  \bibinfo{author}{\bibfnamefont{E.}~\bibnamefont{Kaxiras}}, \bibnamefont{and}
  \bibinfo{author}{\bibfnamefont{P.}~\bibnamefont{Jarillo-Herrero}},
  ``Unconventional superconductivity in magic-angle graphene superlattices,''
  \bibinfo{journal}{Nature} \textbf{\bibinfo{volume}{556}}, \bibinfo{pages}{43}
  (\bibinfo{year}{2018}).

\bibitem[{\citenamefont{Mishra et~al.}(2021{\natexlab{a}})\citenamefont{Mishra,
  Catarina, Wu, Ortiz, Jacob, Eimre, Ma, Pignedoli, Feng, Ruffieux
  et~al.}}]{mishra2021observation}
\bibinfo{author}{\bibfnamefont{S.}~\bibnamefont{Mishra}},
  \bibinfo{author}{\bibfnamefont{G.}~\bibnamefont{Catarina}},
  \bibinfo{author}{\bibfnamefont{F.}~\bibnamefont{Wu}},
  \bibinfo{author}{\bibfnamefont{R.}~\bibnamefont{Ortiz}},
  \bibinfo{author}{\bibfnamefont{D.}~\bibnamefont{Jacob}},
  \bibinfo{author}{\bibfnamefont{K.}~\bibnamefont{Eimre}},
  \bibinfo{author}{\bibfnamefont{J.}~\bibnamefont{Ma}},
  \bibinfo{author}{\bibfnamefont{C.~A.} \bibnamefont{Pignedoli}},
  \bibinfo{author}{\bibfnamefont{X.}~\bibnamefont{Feng}},
  \bibinfo{author}{\bibfnamefont{P.}~\bibnamefont{Ruffieux}},
  \bibnamefont{et~al.}, ``Observation of fractional edge excitations in
  nanographene spin chains,'' \bibinfo{journal}{Nature}
  \textbf{\bibinfo{volume}{598}}, \bibinfo{pages}{287}
  (\bibinfo{year}{2021}{\natexlab{a}}).

\bibitem[{\citenamefont{Mishra et~al.}(2020{\natexlab{a}})\citenamefont{Mishra,
  Beyer, Eimre, Ortiz, Fern{\'a}ndez-Rossier, Berger, Gr{\"o}ning, Pignedoli,
  Fasel, Feng et~al.}}]{mishra2020}
\bibinfo{author}{\bibfnamefont{S.}~\bibnamefont{Mishra}},
  \bibinfo{author}{\bibfnamefont{D.}~\bibnamefont{Beyer}},
  \bibinfo{author}{\bibfnamefont{K.}~\bibnamefont{Eimre}},
  \bibinfo{author}{\bibfnamefont{R.}~\bibnamefont{Ortiz}},
  \bibinfo{author}{\bibfnamefont{J.}~\bibnamefont{Fern{\'a}ndez-Rossier}},
  \bibinfo{author}{\bibfnamefont{R.}~\bibnamefont{Berger}},
  \bibinfo{author}{\bibfnamefont{O.}~\bibnamefont{Gr{\"o}ning}},
  \bibinfo{author}{\bibfnamefont{C.~A.} \bibnamefont{Pignedoli}},
  \bibinfo{author}{\bibfnamefont{R.}~\bibnamefont{Fasel}},
  \bibinfo{author}{\bibfnamefont{X.}~\bibnamefont{Feng}}, \bibnamefont{et~al.},
  ``Collective all-carbon magnetism in triangulene dimers,''
  \bibinfo{journal}{Angewandte Chemie International Edition}
  \textbf{\bibinfo{volume}{59}}, \bibinfo{pages}{12041}
  (\bibinfo{year}{2020}{\natexlab{a}}).

\bibitem[{\citenamefont{Rizzo et~al.}(2020)\citenamefont{Rizzo, Veber, Jiang,
  McCurdy, Cao, Bronner, Chen, Louie, Fischer, and Crommie}}]{crommiescience}
\bibinfo{author}{\bibfnamefont{D.~J.} \bibnamefont{Rizzo}},
  \bibinfo{author}{\bibfnamefont{G.}~\bibnamefont{Veber}},
  \bibinfo{author}{\bibfnamefont{J.}~\bibnamefont{Jiang}},
  \bibinfo{author}{\bibfnamefont{R.}~\bibnamefont{McCurdy}},
  \bibinfo{author}{\bibfnamefont{T.}~\bibnamefont{Cao}},
  \bibinfo{author}{\bibfnamefont{C.}~\bibnamefont{Bronner}},
  \bibinfo{author}{\bibfnamefont{T.}~\bibnamefont{Chen}},
  \bibinfo{author}{\bibfnamefont{S.~G.} \bibnamefont{Louie}},
  \bibinfo{author}{\bibfnamefont{F.~R.} \bibnamefont{Fischer}},
  \bibnamefont{and} \bibinfo{author}{\bibfnamefont{M.~F.}
  \bibnamefont{Crommie}}, ``Inducing metallicity in graphene nanoribbons via
  zero-mode superlattices,'' \bibinfo{journal}{Science}
  \textbf{\bibinfo{volume}{369}}, \bibinfo{pages}{1597} (\bibinfo{year}{2020}).

\bibitem[{\citenamefont{Sun et~al.}(2020{\natexlab{a}})\citenamefont{Sun,
  Gröning, Overbeck, Braun, Perrin, Borin~Barin, El~Abbassi, Eimre, Ditler,
  Daniels et~al.}}]{massivefermions2020}
\bibinfo{author}{\bibfnamefont{Q.}~\bibnamefont{Sun}},
  \bibinfo{author}{\bibfnamefont{O.}~\bibnamefont{Gröning}},
  \bibinfo{author}{\bibfnamefont{J.}~\bibnamefont{Overbeck}},
  \bibinfo{author}{\bibfnamefont{O.}~\bibnamefont{Braun}},
  \bibinfo{author}{\bibfnamefont{M.~L.} \bibnamefont{Perrin}},
  \bibinfo{author}{\bibfnamefont{G.}~\bibnamefont{Borin~Barin}},
  \bibinfo{author}{\bibfnamefont{M.}~\bibnamefont{El~Abbassi}},
  \bibinfo{author}{\bibfnamefont{K.}~\bibnamefont{Eimre}},
  \bibinfo{author}{\bibfnamefont{E.}~\bibnamefont{Ditler}},
  \bibinfo{author}{\bibfnamefont{C.}~\bibnamefont{Daniels}},
  \bibnamefont{et~al.}, ``Massive dirac fermion behavior in a low bandgap
  graphene nanoribbon near a topological phase boundary,''
  \bibinfo{journal}{Advanced Materials} \textbf{\bibinfo{volume}{32}},
  \bibinfo{pages}{1906054} (\bibinfo{year}{2020}{\natexlab{a}}).

\bibitem[{\citenamefont{Sun et~al.}(2020{\natexlab{b}})\citenamefont{Sun, Yao,
  Gr\"oning, Eimre, Pignedoli, M\"ullen, Narita, Fasel, and
  Ruffieux}}]{sun2020coupled}
\bibinfo{author}{\bibfnamefont{Q.}~\bibnamefont{Sun}},
  \bibinfo{author}{\bibfnamefont{X.}~\bibnamefont{Yao}},
  \bibinfo{author}{\bibfnamefont{O.}~\bibnamefont{Gr\"oning}},
  \bibinfo{author}{\bibfnamefont{K.}~\bibnamefont{Eimre}},
  \bibinfo{author}{\bibfnamefont{C.~A.} \bibnamefont{Pignedoli}},
  \bibinfo{author}{\bibfnamefont{K.}~\bibnamefont{M\"ullen}},
  \bibinfo{author}{\bibfnamefont{A.}~\bibnamefont{Narita}},
  \bibinfo{author}{\bibfnamefont{R.}~\bibnamefont{Fasel}}, \bibnamefont{and}
  \bibinfo{author}{\bibfnamefont{P.}~\bibnamefont{Ruffieux}}, ``Coupled spin
  states in armchair graphene nanoribbons with asymmetric zigzag edge
  extensions,'' \bibinfo{journal}{Nano Letters} \textbf{\bibinfo{volume}{20}},
  \bibinfo{pages}{6429} (\bibinfo{year}{2020}{\natexlab{b}}).

\bibitem[{\citenamefont{Sun et~al.}(2021)\citenamefont{Sun, Yan, Yao,
  Mü\"ulen, Narita, Fasel, and Ruffieux}}]{JPCLRuffieux2021}
\bibinfo{author}{\bibfnamefont{Q.}~\bibnamefont{Sun}},
  \bibinfo{author}{\bibfnamefont{Y.}~\bibnamefont{Yan}},
  \bibinfo{author}{\bibfnamefont{X.}~\bibnamefont{Yao}},
  \bibinfo{author}{\bibfnamefont{K.}~\bibnamefont{Mü\"ulen}},
  \bibinfo{author}{\bibfnamefont{A.}~\bibnamefont{Narita}},
  \bibinfo{author}{\bibfnamefont{R.}~\bibnamefont{Fasel}}, \bibnamefont{and}
  \bibinfo{author}{\bibfnamefont{P.}~\bibnamefont{Ruffieux}}, ``Evolution of
  the topological energy band in graphene nanoribbons,'' \bibinfo{journal}{The
  Journal of Physical Chemistry Letters} \textbf{\bibinfo{volume}{12}},
  \bibinfo{pages}{8679} (\bibinfo{year}{2021}).

\bibitem[{\citenamefont{Hieulle et~al.}(2021)\citenamefont{Hieulle, Castro,
  Friedrich, Vegliante, Lara, Sanz, Rey, Corso, Frederiksen, Pascual
  et~al.}}]{NachoNanostar2021}
\bibinfo{author}{\bibfnamefont{J.}~\bibnamefont{Hieulle}},
  \bibinfo{author}{\bibfnamefont{S.}~\bibnamefont{Castro}},
  \bibinfo{author}{\bibfnamefont{N.}~\bibnamefont{Friedrich}},
  \bibinfo{author}{\bibfnamefont{A.}~\bibnamefont{Vegliante}},
  \bibinfo{author}{\bibfnamefont{F.~R.} \bibnamefont{Lara}},
  \bibinfo{author}{\bibfnamefont{S.}~\bibnamefont{Sanz}},
  \bibinfo{author}{\bibfnamefont{D.}~\bibnamefont{Rey}},
  \bibinfo{author}{\bibfnamefont{M.}~\bibnamefont{Corso}},
  \bibinfo{author}{\bibfnamefont{T.}~\bibnamefont{Frederiksen}},
  \bibinfo{author}{\bibfnamefont{J.~I.} \bibnamefont{Pascual}},
  \bibnamefont{et~al.}, ``On-surface synthesis and collective spin excitations
  of a triangulene-based nanostar,'' \bibinfo{journal}{Angewandte Chemie
  International Edition} \textbf{\bibinfo{volume}{60}}, \bibinfo{pages}{25224}
  (\bibinfo{year}{2021}).

\bibitem[{\citenamefont{Gr{\"o}ning et~al.}(2018)\citenamefont{Gr{\"o}ning,
  Wang, Yao, Pignedoli, Borin~Barin, Daniels, Cupo, Meunier, Feng, Narita
  et~al.}}]{groning2018engineering}
\bibinfo{author}{\bibfnamefont{O.}~\bibnamefont{Gr{\"o}ning}},
  \bibinfo{author}{\bibfnamefont{S.}~\bibnamefont{Wang}},
  \bibinfo{author}{\bibfnamefont{X.}~\bibnamefont{Yao}},
  \bibinfo{author}{\bibfnamefont{C.~A.} \bibnamefont{Pignedoli}},
  \bibinfo{author}{\bibfnamefont{G.}~\bibnamefont{Borin~Barin}},
  \bibinfo{author}{\bibfnamefont{C.}~\bibnamefont{Daniels}},
  \bibinfo{author}{\bibfnamefont{A.}~\bibnamefont{Cupo}},
  \bibinfo{author}{\bibfnamefont{V.}~\bibnamefont{Meunier}},
  \bibinfo{author}{\bibfnamefont{X.}~\bibnamefont{Feng}},
  \bibinfo{author}{\bibfnamefont{A.}~\bibnamefont{Narita}},
  \bibnamefont{et~al.}, ``Engineering of robust topological quantum phases in
  graphene nanoribbons,'' \bibinfo{journal}{Nature}
  \textbf{\bibinfo{volume}{560}}, \bibinfo{pages}{209} (\bibinfo{year}{2018}).

\bibitem[{\citenamefont{Ortiz et~al.}(2022{\natexlab{a}})\citenamefont{Ortiz,
  Sancho-Garc\'{\i}a, and Fern\'andez-Rossier}}]{PRMOrtiz2022}
\bibinfo{author}{\bibfnamefont{R.}~\bibnamefont{Ortiz}},
  \bibinfo{author}{\bibfnamefont{J.~C.} \bibnamefont{Sancho-Garc\'{\i}a}},
  \bibnamefont{and}
  \bibinfo{author}{\bibfnamefont{J.}~\bibnamefont{Fern\'andez-Rossier}},
  ``Frustrated magnetic interactions in a cyclacene crystal,''
  \bibinfo{journal}{Phys. Rev. Materials} \textbf{\bibinfo{volume}{6}},
  \bibinfo{pages}{014406} (\bibinfo{year}{2022}{\natexlab{a}}).

\bibitem[{\citenamefont{Catarina and Fern\'andez-Rossier}(2022)}]{toyAKLT}
\bibinfo{author}{\bibfnamefont{G.}~\bibnamefont{Catarina}} \bibnamefont{and}
  \bibinfo{author}{\bibfnamefont{J.}~\bibnamefont{Fern\'andez-Rossier}},
  ``Hubbard model for spin-1 haldane chains,'' \bibinfo{journal}{Phys. Rev. B}
  \textbf{\bibinfo{volume}{105}}, \bibinfo{pages}{L081116}
  (\bibinfo{year}{2022}).

\bibitem[{\citenamefont{Ortiz et~al.}(2022{\natexlab{b}})\citenamefont{Ortiz,
  Catarina, and Fernández-Rossier}}]{2Dtriang}
\bibinfo{author}{\bibfnamefont{R.}~\bibnamefont{Ortiz}},
  \bibinfo{author}{\bibfnamefont{G.}~\bibnamefont{Catarina}}, \bibnamefont{and}
  \bibinfo{author}{\bibfnamefont{J.}~\bibnamefont{Fernández-Rossier}},
  ``Theory of triangulene two-dimensional crystals,''
  (\bibinfo{year}{2022}{\natexlab{b}}).

\bibitem[{\citenamefont{Wang et~al.}(2016)\citenamefont{Wang, Talirz,
  Pignedoli, Feng, M{\"u}llen, Fasel, and Ruffieux}}]{wang2016}
\bibinfo{author}{\bibfnamefont{S.}~\bibnamefont{Wang}},
  \bibinfo{author}{\bibfnamefont{L.}~\bibnamefont{Talirz}},
  \bibinfo{author}{\bibfnamefont{C.~A.} \bibnamefont{Pignedoli}},
  \bibinfo{author}{\bibfnamefont{X.}~\bibnamefont{Feng}},
  \bibinfo{author}{\bibfnamefont{K.}~\bibnamefont{M{\"u}llen}},
  \bibinfo{author}{\bibfnamefont{R.}~\bibnamefont{Fasel}}, \bibnamefont{and}
  \bibinfo{author}{\bibfnamefont{P.}~\bibnamefont{Ruffieux}}, ``Giant edge
  state splitting at atomically precise graphene zigzag edges,''
  \bibinfo{journal}{Nature Communications} \textbf{\bibinfo{volume}{7}},
  \bibinfo{pages}{11507} (\bibinfo{year}{2016}).

\bibitem[{\citenamefont{Ruffieux et~al.}(2016)\citenamefont{Ruffieux, Wang,
  Yang, S{\'a}nchez-S{\'a}nchez, Liu, Dienel, Talirz, Shinde, Pignedoli,
  Passerone et~al.}}]{Ruffieux16}
\bibinfo{author}{\bibfnamefont{P.}~\bibnamefont{Ruffieux}},
  \bibinfo{author}{\bibfnamefont{S.}~\bibnamefont{Wang}},
  \bibinfo{author}{\bibfnamefont{B.}~\bibnamefont{Yang}},
  \bibinfo{author}{\bibfnamefont{C.}~\bibnamefont{S{\'a}nchez-S{\'a}nchez}},
  \bibinfo{author}{\bibfnamefont{J.}~\bibnamefont{Liu}},
  \bibinfo{author}{\bibfnamefont{T.}~\bibnamefont{Dienel}},
  \bibinfo{author}{\bibfnamefont{L.}~\bibnamefont{Talirz}},
  \bibinfo{author}{\bibfnamefont{P.}~\bibnamefont{Shinde}},
  \bibinfo{author}{\bibfnamefont{C.~A.} \bibnamefont{Pignedoli}},
  \bibinfo{author}{\bibfnamefont{D.}~\bibnamefont{Passerone}},
  \bibnamefont{et~al.}, ``On-surface synthesis of graphene nanoribbons with
  zigzag edge topology,'' \bibinfo{journal}{Nature}
  \textbf{\bibinfo{volume}{531}}, \bibinfo{pages}{489} (\bibinfo{year}{2016}).

\bibitem[{\citenamefont{Pavli{\v{c}}ek
  et~al.}(2017)\citenamefont{Pavli{\v{c}}ek, Mistry, Majzik, Moll, Meyer, Fox,
  and Gross}}]{pavlivcek2017synthesis}
\bibinfo{author}{\bibfnamefont{N.}~\bibnamefont{Pavli{\v{c}}ek}},
  \bibinfo{author}{\bibfnamefont{A.}~\bibnamefont{Mistry}},
  \bibinfo{author}{\bibfnamefont{Z.}~\bibnamefont{Majzik}},
  \bibinfo{author}{\bibfnamefont{N.}~\bibnamefont{Moll}},
  \bibinfo{author}{\bibfnamefont{G.}~\bibnamefont{Meyer}},
  \bibinfo{author}{\bibfnamefont{D.~J.} \bibnamefont{Fox}}, \bibnamefont{and}
  \bibinfo{author}{\bibfnamefont{L.}~\bibnamefont{Gross}}, ``Synthesis and
  characterization of triangulene,'' \bibinfo{journal}{Nature Nanotechnology}
  \textbf{\bibinfo{volume}{12}}, \bibinfo{pages}{308} (\bibinfo{year}{2017}).

\bibitem[{\citenamefont{Mishra et~al.}(2018)\citenamefont{Mishra, Lohr,
  Pignedoli, Liu, Berger, Urgel, Müllen, Feng, Ruffieux, and
  Fasel}}]{tailoringbondMishra2018}
\bibinfo{author}{\bibfnamefont{S.}~\bibnamefont{Mishra}},
  \bibinfo{author}{\bibfnamefont{T.~G.} \bibnamefont{Lohr}},
  \bibinfo{author}{\bibfnamefont{C.~A.} \bibnamefont{Pignedoli}},
  \bibinfo{author}{\bibfnamefont{J.}~\bibnamefont{Liu}},
  \bibinfo{author}{\bibfnamefont{R.}~\bibnamefont{Berger}},
  \bibinfo{author}{\bibfnamefont{J.~I.} \bibnamefont{Urgel}},
  \bibinfo{author}{\bibfnamefont{K.}~\bibnamefont{Müllen}},
  \bibinfo{author}{\bibfnamefont{X.}~\bibnamefont{Feng}},
  \bibinfo{author}{\bibfnamefont{P.}~\bibnamefont{Ruffieux}}, \bibnamefont{and}
  \bibinfo{author}{\bibfnamefont{R.}~\bibnamefont{Fasel}}, ``Tailoring bond
  topologies in open-shell graphene nanostructures,'' \bibinfo{journal}{ACS
  Nano} \textbf{\bibinfo{volume}{12}}, \bibinfo{pages}{11917}
  (\bibinfo{year}{2018}).

\bibitem[{\citenamefont{Li et~al.}(2019)\citenamefont{Li, Sanz, Corso, Choi,
  Pe{\~n}a, Frederiksen, and Pascual}}]{NachoNat}
\bibinfo{author}{\bibfnamefont{J.}~\bibnamefont{Li}},
  \bibinfo{author}{\bibfnamefont{S.}~\bibnamefont{Sanz}},
  \bibinfo{author}{\bibfnamefont{M.}~\bibnamefont{Corso}},
  \bibinfo{author}{\bibfnamefont{D.~J.} \bibnamefont{Choi}},
  \bibinfo{author}{\bibfnamefont{D.}~\bibnamefont{Pe{\~n}a}},
  \bibinfo{author}{\bibfnamefont{T.}~\bibnamefont{Frederiksen}},
  \bibnamefont{and} \bibinfo{author}{\bibfnamefont{J.~I.}
  \bibnamefont{Pascual}}, ``Single spin localization and manipulation in
  graphene open-shell nanostructures,'' \bibinfo{journal}{Nature
  Communications} \textbf{\bibinfo{volume}{10}} (\bibinfo{year}{2019}).

\bibitem[{\citenamefont{Li et~al.}(2020)\citenamefont{Li, Sanz, Castro-Esteban,
  Vilas-Varela, Friedrich, Frederiksen, Pe\~na, and Pascual}}]{li2019}
\bibinfo{author}{\bibfnamefont{J.}~\bibnamefont{Li}},
  \bibinfo{author}{\bibfnamefont{S.}~\bibnamefont{Sanz}},
  \bibinfo{author}{\bibfnamefont{J.}~\bibnamefont{Castro-Esteban}},
  \bibinfo{author}{\bibfnamefont{M.}~\bibnamefont{Vilas-Varela}},
  \bibinfo{author}{\bibfnamefont{N.}~\bibnamefont{Friedrich}},
  \bibinfo{author}{\bibfnamefont{T.}~\bibnamefont{Frederiksen}},
  \bibinfo{author}{\bibfnamefont{D.}~\bibnamefont{Pe\~na}}, \bibnamefont{and}
  \bibinfo{author}{\bibfnamefont{J.~I.} \bibnamefont{Pascual}}, ``Uncovering
  the triplet ground state of triangular graphene nanoflakes engineered with
  atomic precision on a metal surface,'' \bibinfo{journal}{Physical Review
  Letters} \textbf{\bibinfo{volume}{124}}, \bibinfo{pages}{177201}
  (\bibinfo{year}{2020}).

\bibitem[{\citenamefont{Mishra et~al.}(2020{\natexlab{b}})\citenamefont{Mishra,
  Beyer, Berger, Liu, Gröning, Urgel, Müllen, Ruffieux, Feng, and
  Fasel}}]{ShantaJACS20}
\bibinfo{author}{\bibfnamefont{S.}~\bibnamefont{Mishra}},
  \bibinfo{author}{\bibfnamefont{D.}~\bibnamefont{Beyer}},
  \bibinfo{author}{\bibfnamefont{R.}~\bibnamefont{Berger}},
  \bibinfo{author}{\bibfnamefont{J.}~\bibnamefont{Liu}},
  \bibinfo{author}{\bibfnamefont{O.}~\bibnamefont{Gröning}},
  \bibinfo{author}{\bibfnamefont{J.~I.} \bibnamefont{Urgel}},
  \bibinfo{author}{\bibfnamefont{K.}~\bibnamefont{Müllen}},
  \bibinfo{author}{\bibfnamefont{P.}~\bibnamefont{Ruffieux}},
  \bibinfo{author}{\bibfnamefont{X.}~\bibnamefont{Feng}}, \bibnamefont{and}
  \bibinfo{author}{\bibfnamefont{R.}~\bibnamefont{Fasel}}, ``Topological
  defect-induced magnetism in a nanographene,'' \bibinfo{journal}{Journal of
  the American Chemical Society} \textbf{\bibinfo{volume}{142}},
  \bibinfo{pages}{1147} (\bibinfo{year}{2020}{\natexlab{b}}).

\bibitem[{\citenamefont{Mishra et~al.}(2021{\natexlab{b}})\citenamefont{Mishra,
  Yao, Chen, Eimre, Gr{\"o}ning, Ortiz, Di-Giovannantonio, Sancho-Garc\'ia,
  Fern\'andez-Rossier, Pignedoli et~al.}}]{rhombenes}
\bibinfo{author}{\bibfnamefont{S.}~\bibnamefont{Mishra}},
  \bibinfo{author}{\bibfnamefont{X.}~\bibnamefont{Yao}},
  \bibinfo{author}{\bibfnamefont{Q.}~\bibnamefont{Chen}},
  \bibinfo{author}{\bibfnamefont{K.}~\bibnamefont{Eimre}},
  \bibinfo{author}{\bibfnamefont{O.}~\bibnamefont{Gr{\"o}ning}},
  \bibinfo{author}{\bibfnamefont{R.}~\bibnamefont{Ortiz}},
  \bibinfo{author}{\bibfnamefont{M.}~\bibnamefont{Di-Giovannantonio}},
  \bibinfo{author}{\bibfnamefont{J.~C.} \bibnamefont{Sancho-Garc\'ia}},
  \bibinfo{author}{\bibfnamefont{J.}~\bibnamefont{Fern\'andez-Rossier}},
  \bibinfo{author}{\bibfnamefont{C.~A.} \bibnamefont{Pignedoli}},
  \bibnamefont{et~al.}, ``Large magnetic exchange coupling in rhombus-shaped
  nanographenes with zigzag periphery,'' \bibinfo{journal}{Nature Chemistry}
  \textbf{\bibinfo{volume}{13}}, \bibinfo{pages}{581}
  (\bibinfo{year}{2021}{\natexlab{b}}).

\bibitem[{\citenamefont{Mishra et~al.}(2019)\citenamefont{Mishra, Beyer, Eimre,
  Kezilebieke, Berger, Gr{\"o}ning, Pignedoli, M{\"u}llen, Liljeroth, Ruffieux
  et~al.}}]{mishra19b}
\bibinfo{author}{\bibfnamefont{S.}~\bibnamefont{Mishra}},
  \bibinfo{author}{\bibfnamefont{D.}~\bibnamefont{Beyer}},
  \bibinfo{author}{\bibfnamefont{K.}~\bibnamefont{Eimre}},
  \bibinfo{author}{\bibfnamefont{S.}~\bibnamefont{Kezilebieke}},
  \bibinfo{author}{\bibfnamefont{R.}~\bibnamefont{Berger}},
  \bibinfo{author}{\bibfnamefont{O.}~\bibnamefont{Gr{\"o}ning}},
  \bibinfo{author}{\bibfnamefont{C.~A.} \bibnamefont{Pignedoli}},
  \bibinfo{author}{\bibfnamefont{K.}~\bibnamefont{M{\"u}llen}},
  \bibinfo{author}{\bibfnamefont{P.}~\bibnamefont{Liljeroth}},
  \bibinfo{author}{\bibfnamefont{P.}~\bibnamefont{Ruffieux}},
  \bibnamefont{et~al.}, ``Topological frustration induces unconventional
  magnetism in a nanographene,'' \bibinfo{journal}{Nature Nanotechnology} pp.
  \bibinfo{pages}{1--7} (\bibinfo{year}{2019}).

\bibitem[{\citenamefont{Ortiz and Fernández-Rossier}(2020)}]{ORTIZ2020100595}
\bibinfo{author}{\bibfnamefont{R.}~\bibnamefont{Ortiz}} \bibnamefont{and}
  \bibinfo{author}{\bibfnamefont{J.}~\bibnamefont{Fernández-Rossier}},
  ``Probing local moments in nanographenes with electron tunneling
  spectroscopy,'' \bibinfo{journal}{Progress in Surface Science}
  \textbf{\bibinfo{volume}{95}}, \bibinfo{pages}{100595}
  (\bibinfo{year}{2020}).

\bibitem[{\citenamefont{Song et~al.}(2021)\citenamefont{Song, Su, Telychko, Li,
  Li, Li, Su, Wu, and Lu}}]{D0CS01060J}
\bibinfo{author}{\bibfnamefont{S.}~\bibnamefont{Song}},
  \bibinfo{author}{\bibfnamefont{J.}~\bibnamefont{Su}},
  \bibinfo{author}{\bibfnamefont{M.}~\bibnamefont{Telychko}},
  \bibinfo{author}{\bibfnamefont{J.}~\bibnamefont{Li}},
  \bibinfo{author}{\bibfnamefont{G.}~\bibnamefont{Li}},
  \bibinfo{author}{\bibfnamefont{Y.}~\bibnamefont{Li}},
  \bibinfo{author}{\bibfnamefont{C.}~\bibnamefont{Su}},
  \bibinfo{author}{\bibfnamefont{J.}~\bibnamefont{Wu}}, \bibnamefont{and}
  \bibinfo{author}{\bibfnamefont{J.}~\bibnamefont{Lu}}, ``On-surface synthesis
  of graphene nanostructures with $\pi$-magnetism,'' \bibinfo{journal}{Chem.
  Soc. Rev.} \textbf{\bibinfo{volume}{50}}, \bibinfo{pages}{3238}
  (\bibinfo{year}{2021}).

\bibitem[{\citenamefont{de~Oteyza et~al.}(2022)\citenamefont{de~Oteyza, Dimas,
  and Frederiksen}}]{OtDiFr.22.Carbonbasednanostructures}
\bibinfo{author}{\bibnamefont{de~Oteyza}},
  \bibinfo{author}{\bibfnamefont{G.}~\bibnamefont{Dimas}}, \bibnamefont{and}
  \bibinfo{author}{\bibfnamefont{T.}~\bibnamefont{Frederiksen}}, ``Carbon-based
  nanostructures as a versatile platform for tunable $\pi$-magnetism,''
  \bibinfo{journal}{J. Phys.: Condens. Matter} \textbf{\bibinfo{volume}{34}},
  \bibinfo{pages}{443001} (\bibinfo{year}{2022}).

\bibitem[{\citenamefont{Lado et~al.}(2015)\citenamefont{Lado,
  García-Martínez, and Fernández-Rossier}}]{LADO201556}
\bibinfo{author}{\bibfnamefont{J.}~\bibnamefont{Lado}},
  \bibinfo{author}{\bibfnamefont{N.}~\bibnamefont{García-Martínez}},
  \bibnamefont{and}
  \bibinfo{author}{\bibfnamefont{J.}~\bibnamefont{Fernández-Rossier}}, ``Edge
  states in graphene-like systems,'' \bibinfo{journal}{Synthetic Metals}
  \textbf{\bibinfo{volume}{210}}, \bibinfo{pages}{56} (\bibinfo{year}{2015}),
  \bibinfo{note}{reviews of Current Advances in Graphene Science and
  Technology}.

\bibitem[{\citenamefont{Fern{\'a}ndez-Rossier and Palacios}(2007)}]{JFR07}
\bibinfo{author}{\bibfnamefont{J.}~\bibnamefont{Fern{\'a}ndez-Rossier}}
  \bibnamefont{and} \bibinfo{author}{\bibfnamefont{J.~J.}
  \bibnamefont{Palacios}}, ``Magnetism in graphene nanoislands,''
  \bibinfo{journal}{Physical Review Letters} \textbf{\bibinfo{volume}{99}},
  \bibinfo{pages}{177204} (\bibinfo{year}{2007}).

\bibitem[{\citenamefont{Koshino et~al.}(2014)\citenamefont{Koshino, Morimoto,
  and Sato}}]{koshino2014topological}
\bibinfo{author}{\bibfnamefont{M.}~\bibnamefont{Koshino}},
  \bibinfo{author}{\bibfnamefont{T.}~\bibnamefont{Morimoto}}, \bibnamefont{and}
  \bibinfo{author}{\bibfnamefont{M.}~\bibnamefont{Sato}}, ``Topological zero
  modes and dirac points protected by spatial symmetry and chiral symmetry,''
  \bibinfo{journal}{Physical Review B} \textbf{\bibinfo{volume}{90}},
  \bibinfo{pages}{115207} (\bibinfo{year}{2014}).

\bibitem[{\citenamefont{Cao et~al.}(2017)\citenamefont{Cao, Zhao, and
  Louie}}]{louiejunction}
\bibinfo{author}{\bibfnamefont{T.}~\bibnamefont{Cao}},
  \bibinfo{author}{\bibfnamefont{F.}~\bibnamefont{Zhao}}, \bibnamefont{and}
  \bibinfo{author}{\bibfnamefont{S.~G.} \bibnamefont{Louie}}, ``Topological
  phases in graphene nanoribbons: Junction states, spin centers, and quantum
  spin chains,'' \bibinfo{journal}{Phys. Rev. Lett.}
  \textbf{\bibinfo{volume}{119}}, \bibinfo{pages}{076401}
  (\bibinfo{year}{2017}).

\bibitem[{\citenamefont{Ortiz et~al.}(2019)\citenamefont{Ortiz, Boto,
  Garc{\'\i}a-Mart{\'\i}nez, Sancho-Garc{\'\i}a, Melle-Franco, and
  Fern{\'a}ndez-Rossier}}]{ortiz19}
\bibinfo{author}{\bibfnamefont{R.}~\bibnamefont{Ortiz}},
  \bibinfo{author}{\bibfnamefont{R.~{\'A}.} \bibnamefont{Boto}},
  \bibinfo{author}{\bibfnamefont{N.}~\bibnamefont{Garc{\'\i}a-Mart{\'\i}nez}},
  \bibinfo{author}{\bibfnamefont{J.~C.} \bibnamefont{Sancho-Garc{\'\i}a}},
  \bibinfo{author}{\bibfnamefont{M.}~\bibnamefont{Melle-Franco}},
  \bibnamefont{and}
  \bibinfo{author}{\bibfnamefont{J.}~\bibnamefont{Fern{\'a}ndez-Rossier}},
  ``Exchange rules for diradical $\pi$-conjugated hydrocarbons,''
  \bibinfo{journal}{Nano Letters} \textbf{\bibinfo{volume}{19}},
  \bibinfo{pages}{5991} (\bibinfo{year}{2019}).

\bibitem[{\citenamefont{Jacob and
  Fernández-Rossier}(2022)}]{superexchangejacob}
\bibinfo{author}{\bibfnamefont{D.}~\bibnamefont{Jacob}} \bibnamefont{and}
  \bibinfo{author}{\bibfnamefont{J.}~\bibnamefont{Fernández-Rossier}},
  ``Theory of intermolecular exchange in coupled spin-1/2 nanographenes,''
  (\bibinfo{year}{2022}).

\bibitem[{\citenamefont{Lieb}(1989)}]{Lieb89}
\bibinfo{author}{\bibfnamefont{E.~H.} \bibnamefont{Lieb}}, ``Two theorems on
  the hubbard model,'' \bibinfo{journal}{Physical Review Letters}
  \textbf{\bibinfo{volume}{62}}, \bibinfo{pages}{1201} (\bibinfo{year}{1989}).

\bibitem[{\citenamefont{Cusinato et~al.}(2018)\citenamefont{Cusinato,
  Evangelisti, Leininger, and Monari}}]{cusinato2018electronic}
\bibinfo{author}{\bibfnamefont{L.}~\bibnamefont{Cusinato}},
  \bibinfo{author}{\bibfnamefont{S.}~\bibnamefont{Evangelisti}},
  \bibinfo{author}{\bibfnamefont{T.}~\bibnamefont{Leininger}},
  \bibnamefont{and} \bibinfo{author}{\bibfnamefont{A.}~\bibnamefont{Monari}},
  ``The electronic structure of graphene nanoislands: A cas-scf and nevpt2
  study,'' \bibinfo{journal}{Advances in Condensed Matter Physics}
  \textbf{\bibinfo{volume}{2018}} (\bibinfo{year}{2018}).

\bibitem[{\citenamefont{Ortiz et~al.}(2016)\citenamefont{Ortiz, Lado,
  Melle-Franco, and Fern\'andez-Rossier}}]{graphsatelites}
\bibinfo{author}{\bibfnamefont{R.}~\bibnamefont{Ortiz}},
  \bibinfo{author}{\bibfnamefont{J.~L.} \bibnamefont{Lado}},
  \bibinfo{author}{\bibfnamefont{M.}~\bibnamefont{Melle-Franco}},
  \bibnamefont{and}
  \bibinfo{author}{\bibfnamefont{J.}~\bibnamefont{Fern\'andez-Rossier}},
  ``Engineering spin exchange in nonbipartite graphene zigzag edges,''
  \bibinfo{journal}{Phys. Rev. B} \textbf{\bibinfo{volume}{94}},
  \bibinfo{pages}{094414} (\bibinfo{year}{2016}).

\bibitem[{\citenamefont{Anderson}(1959)}]{andersonexchange}
\bibinfo{author}{\bibfnamefont{P.~W.} \bibnamefont{Anderson}}, ``New approach
  to the theory of superexchange interactions,'' \bibinfo{journal}{Phys. Rev.}
  \textbf{\bibinfo{volume}{115}}, \bibinfo{pages}{2} (\bibinfo{year}{1959}).

\bibitem[{\citenamefont{Ortiz et~al.}(2018)\citenamefont{Ortiz,
  Garc\'{\i}a-Mart\'{\i}nez, Lado, and Fern\'andez-Rossier}}]{Ortiz18}
\bibinfo{author}{\bibfnamefont{R.}~\bibnamefont{Ortiz}},
  \bibinfo{author}{\bibfnamefont{N.~A.}
  \bibnamefont{Garc\'{\i}a-Mart\'{\i}nez}},
  \bibinfo{author}{\bibfnamefont{J.~L.} \bibnamefont{Lado}}, \bibnamefont{and}
  \bibinfo{author}{\bibfnamefont{J.}~\bibnamefont{Fern\'andez-Rossier}},
  ``Electrical spin manipulation in graphene nanostructures,''
  \bibinfo{journal}{Physical Review B} \textbf{\bibinfo{volume}{97}},
  \bibinfo{pages}{195425} (\bibinfo{year}{2018}).

\bibitem[{\citenamefont{Majzik et~al.}(2018)\citenamefont{Majzik,
  Pavli{\v{c}}ek, Vilas-Varela, P\'erez, Moll, Guiti\'an, Meyer, Peña, and
  Gross}}]{transdimerexp}
\bibinfo{author}{\bibfnamefont{Z.}~\bibnamefont{Majzik}},
  \bibinfo{author}{\bibfnamefont{N.}~\bibnamefont{Pavli{\v{c}}ek}},
  \bibinfo{author}{\bibfnamefont{M.}~\bibnamefont{Vilas-Varela}},
  \bibinfo{author}{\bibfnamefont{D.}~\bibnamefont{P\'erez}},
  \bibinfo{author}{\bibfnamefont{N.}~\bibnamefont{Moll}},
  \bibinfo{author}{\bibfnamefont{E.}~\bibnamefont{Guiti\'an}},
  \bibinfo{author}{\bibfnamefont{G.}~\bibnamefont{Meyer}},
  \bibinfo{author}{\bibfnamefont{D.}~\bibnamefont{Peña}}, \bibnamefont{and}
  \bibinfo{author}{\bibfnamefont{L.}~\bibnamefont{Gross}}, ``Studying an
  antiaromatic polycyclic hydrocarbon adsorbed on different surfaces,''
  \bibinfo{journal}{Nature Communications} \textbf{\bibinfo{volume}{9}}
  (\bibinfo{year}{2018}).

\bibitem[{\citenamefont{Di~Giovannantonio
  et~al.}(2020)\citenamefont{Di~Giovannantonio, Chen, Urgel, Ruffieux,
  Pignedoli, M\"ullen, Narita, and Fasel}}]{transchainexp}
\bibinfo{author}{\bibfnamefont{M.}~\bibnamefont{Di~Giovannantonio}},
  \bibinfo{author}{\bibfnamefont{Q.}~\bibnamefont{Chen}},
  \bibinfo{author}{\bibfnamefont{J.~I.} \bibnamefont{Urgel}},
  \bibinfo{author}{\bibfnamefont{P.}~\bibnamefont{Ruffieux}},
  \bibinfo{author}{\bibfnamefont{C.~A.} \bibnamefont{Pignedoli}},
  \bibinfo{author}{\bibfnamefont{K.}~\bibnamefont{M\"ullen}},
  \bibinfo{author}{\bibfnamefont{A.}~\bibnamefont{Narita}}, \bibnamefont{and}
  \bibinfo{author}{\bibfnamefont{R.}~\bibnamefont{Fasel}}, ``On-surface
  synthesis of oligo(indenoindene),'' \bibinfo{journal}{Journal of the American
  Chemical Society} \textbf{\bibinfo{volume}{142}}, \bibinfo{pages}{12925}
  (\bibinfo{year}{2020}).

\bibitem[{\citenamefont{Fukuda et~al.}(2015)\citenamefont{Fukuda, Nagami,
  Fujiyoshi, and Nakano}}]{JACSN2}
\bibinfo{author}{\bibfnamefont{K.}~\bibnamefont{Fukuda}},
  \bibinfo{author}{\bibfnamefont{T.}~\bibnamefont{Nagami}},
  \bibinfo{author}{\bibfnamefont{J.-y.} \bibnamefont{Fujiyoshi}},
  \bibnamefont{and} \bibinfo{author}{\bibfnamefont{M.}~\bibnamefont{Nakano}},
  ``Interplay between open-shell character, aromaticity, and second
  hyperpolarizabilities in indenofluorenes,'' \bibinfo{journal}{The Journal of
  Physical Chemistry A} \textbf{\bibinfo{volume}{119}}, \bibinfo{pages}{10620}
  (\bibinfo{year}{2015}).

\bibitem[{\citenamefont{Thomas and Kim}(2014)}]{C4CP03169E}
\bibinfo{author}{\bibfnamefont{S.}~\bibnamefont{Thomas}} \bibnamefont{and}
  \bibinfo{author}{\bibfnamefont{K.~S.} \bibnamefont{Kim}}, ``Linear and
  nonlinear optical properties of indeno[2,1-b]fluorene and its structural
  isomers,'' \bibinfo{journal}{Phys. Chem. Chem. Phys.}
  \textbf{\bibinfo{volume}{16}}, \bibinfo{pages}{24592} (\bibinfo{year}{2014}).

\bibitem[{\citenamefont{Yamane et~al.}(2018)\citenamefont{Yamane, Kishi,
  Tonami, Okada, Nagami, Kitagawa, and Nakano}}]{yamane2018open}
\bibinfo{author}{\bibfnamefont{M.}~\bibnamefont{Yamane}},
  \bibinfo{author}{\bibfnamefont{R.}~\bibnamefont{Kishi}},
  \bibinfo{author}{\bibfnamefont{T.}~\bibnamefont{Tonami}},
  \bibinfo{author}{\bibfnamefont{K.}~\bibnamefont{Okada}},
  \bibinfo{author}{\bibfnamefont{T.}~\bibnamefont{Nagami}},
  \bibinfo{author}{\bibfnamefont{Y.}~\bibnamefont{Kitagawa}}, \bibnamefont{and}
  \bibinfo{author}{\bibfnamefont{M.}~\bibnamefont{Nakano}}, ``Open-shell
  characters, aromaticities and third-order nonlinear optical properties of
  carbon nanobelts composed of five-and six-membered rings,''
  \bibinfo{journal}{Asian Journal of Organic Chemistry}
  \textbf{\bibinfo{volume}{7}}, \bibinfo{pages}{2320} (\bibinfo{year}{2018}).

\bibitem[{\citenamefont{Fukuda et~al.}(2017)\citenamefont{Fukuda, Fujiyoshi,
  Matsui, Nagami, Takamuku, Kitagawa, Champagne, and
  Nakano}}]{fukuda2017theoretical}
\bibinfo{author}{\bibfnamefont{K.}~\bibnamefont{Fukuda}},
  \bibinfo{author}{\bibfnamefont{J.-y.} \bibnamefont{Fujiyoshi}},
  \bibinfo{author}{\bibfnamefont{H.}~\bibnamefont{Matsui}},
  \bibinfo{author}{\bibfnamefont{T.}~\bibnamefont{Nagami}},
  \bibinfo{author}{\bibfnamefont{S.}~\bibnamefont{Takamuku}},
  \bibinfo{author}{\bibfnamefont{Y.}~\bibnamefont{Kitagawa}},
  \bibinfo{author}{\bibfnamefont{B.}~\bibnamefont{Champagne}},
  \bibnamefont{and} \bibinfo{author}{\bibfnamefont{M.}~\bibnamefont{Nakano}},
  ``A theoretical study on quasi-one-dimensional open-shell singlet ladder
  oligomers: multi-radical nature, aromaticity and second
  hyperpolarizability,'' \bibinfo{journal}{Organic Chemistry Frontiers}
  \textbf{\bibinfo{volume}{4}}, \bibinfo{pages}{779} (\bibinfo{year}{2017}).

\bibitem[{\citenamefont{Mishra et~al.}(2022)\citenamefont{Mishra, Fatayer,
  Fernández, Kaiser, Peña, and Gross}}]{truxene}
\bibinfo{author}{\bibfnamefont{S.}~\bibnamefont{Mishra}},
  \bibinfo{author}{\bibfnamefont{S.}~\bibnamefont{Fatayer}},
  \bibinfo{author}{\bibfnamefont{S.}~\bibnamefont{Fernández}},
  \bibinfo{author}{\bibfnamefont{K.}~\bibnamefont{Kaiser}},
  \bibinfo{author}{\bibfnamefont{D.}~\bibnamefont{Peña}}, \bibnamefont{and}
  \bibinfo{author}{\bibfnamefont{L.}~\bibnamefont{Gross}}, ``Nonbenzenoid
  high-spin polycyclic hydrocarbons generated by atom manipulation,''
  \bibinfo{journal}{ACS Nano} \textbf{\bibinfo{volume}{16}},
  \bibinfo{pages}{3264} (\bibinfo{year}{2022}).

\bibitem[{\citenamefont{Li et~al.}(2022)\citenamefont{Li, Liu, Liu, Xue, Guan,
  Li, Zheng, Liu, Jia, Liu et~al.}}]{truxenewang}
\bibinfo{author}{\bibfnamefont{C.}~\bibnamefont{Li}},
  \bibinfo{author}{\bibfnamefont{Y.}~\bibnamefont{Liu}},
  \bibinfo{author}{\bibfnamefont{Y.}~\bibnamefont{Liu}},
  \bibinfo{author}{\bibfnamefont{F.-H.} \bibnamefont{Xue}},
  \bibinfo{author}{\bibfnamefont{D.}~\bibnamefont{Guan}},
  \bibinfo{author}{\bibfnamefont{Y.}~\bibnamefont{Li}},
  \bibinfo{author}{\bibfnamefont{H.}~\bibnamefont{Zheng}},
  \bibinfo{author}{\bibfnamefont{C.}~\bibnamefont{Liu}},
  \bibinfo{author}{\bibfnamefont{J.}~\bibnamefont{Jia}},
  \bibinfo{author}{\bibfnamefont{P.-N.} \bibnamefont{Liu}},
  \bibnamefont{et~al.}, ``Topological defects induced high-spin quartet state
  in truxene-based molecular graphenoids,'' \bibinfo{journal}{CCS Chemistry}
  \textbf{\bibinfo{volume}{0}}, \bibinfo{pages}{1} (\bibinfo{year}{2022}).

\bibitem[{\citenamefont{Di~Giovannantonio
  et~al.}(2019)\citenamefont{Di~Giovannantonio, Eimre, Yakutovich, Chen,
  Mishra, Urgel, Pignedoli, Ruffieux, Müllen, Narita et~al.}}]{chainisomer5}
\bibinfo{author}{\bibfnamefont{M.}~\bibnamefont{Di~Giovannantonio}},
  \bibinfo{author}{\bibfnamefont{K.}~\bibnamefont{Eimre}},
  \bibinfo{author}{\bibfnamefont{A.~V.} \bibnamefont{Yakutovich}},
  \bibinfo{author}{\bibfnamefont{Q.}~\bibnamefont{Chen}},
  \bibinfo{author}{\bibfnamefont{S.}~\bibnamefont{Mishra}},
  \bibinfo{author}{\bibfnamefont{J.~I.} \bibnamefont{Urgel}},
  \bibinfo{author}{\bibfnamefont{C.~A.} \bibnamefont{Pignedoli}},
  \bibinfo{author}{\bibfnamefont{P.}~\bibnamefont{Ruffieux}},
  \bibinfo{author}{\bibfnamefont{K.}~\bibnamefont{Müllen}},
  \bibinfo{author}{\bibfnamefont{A.}~\bibnamefont{Narita}},
  \bibnamefont{et~al.}, ``On-surface synthesis of antiaromatic and open-shell
  indeno[2,1-b]fluorene polymers and their lateral fusion into porous
  ribbons,'' \bibinfo{journal}{Journal of the American Chemical Society}
  \textbf{\bibinfo{volume}{141}}, \bibinfo{pages}{12346}
  (\bibinfo{year}{2019}).

\bibitem[{\citenamefont{Di~Giovannantonio
  et~al.}(2018)\citenamefont{Di~Giovannantonio, Urgel, Beser, Yakutovich,
  Wilhelm, Pignedoli, Ruffieux, Narita, Müllen, and
  Fasel}}]{intercalationsJACS}
\bibinfo{author}{\bibfnamefont{M.}~\bibnamefont{Di~Giovannantonio}},
  \bibinfo{author}{\bibfnamefont{J.~I.} \bibnamefont{Urgel}},
  \bibinfo{author}{\bibfnamefont{U.}~\bibnamefont{Beser}},
  \bibinfo{author}{\bibfnamefont{A.~V.} \bibnamefont{Yakutovich}},
  \bibinfo{author}{\bibfnamefont{J.}~\bibnamefont{Wilhelm}},
  \bibinfo{author}{\bibfnamefont{C.~A.} \bibnamefont{Pignedoli}},
  \bibinfo{author}{\bibfnamefont{P.}~\bibnamefont{Ruffieux}},
  \bibinfo{author}{\bibfnamefont{A.}~\bibnamefont{Narita}},
  \bibinfo{author}{\bibfnamefont{K.}~\bibnamefont{Müllen}}, \bibnamefont{and}
  \bibinfo{author}{\bibfnamefont{R.}~\bibnamefont{Fasel}}, ``On-surface
  synthesis of indenofluorene polymers by oxidative five-membered ring
  formation,'' \bibinfo{journal}{Journal of the American Chemical Society}
  \textbf{\bibinfo{volume}{140}}, \bibinfo{pages}{3532} (\bibinfo{year}{2018}).

\bibitem[{\citenamefont{Hu et~al.}(2016)\citenamefont{Hu, Lee, Herng, Aratani,
  Gon{\c{c}}alves, Qi, Shi, Yamada, Huang, Ding et~al.}}]{hu2016toward}
\bibinfo{author}{\bibfnamefont{P.}~\bibnamefont{Hu}},
  \bibinfo{author}{\bibfnamefont{S.}~\bibnamefont{Lee}},
  \bibinfo{author}{\bibfnamefont{T.~S.} \bibnamefont{Herng}},
  \bibinfo{author}{\bibfnamefont{N.}~\bibnamefont{Aratani}},
  \bibinfo{author}{\bibfnamefont{T.~P.} \bibnamefont{Gon{\c{c}}alves}},
  \bibinfo{author}{\bibfnamefont{Q.}~\bibnamefont{Qi}},
  \bibinfo{author}{\bibfnamefont{X.}~\bibnamefont{Shi}},
  \bibinfo{author}{\bibfnamefont{H.}~\bibnamefont{Yamada}},
  \bibinfo{author}{\bibfnamefont{K.-W.} \bibnamefont{Huang}},
  \bibinfo{author}{\bibfnamefont{J.}~\bibnamefont{Ding}}, \bibnamefont{et~al.},
  ``Toward tetraradicaloid: the effect of fusion mode on radical character and
  chemical reactivity,'' \bibinfo{journal}{Journal of the American Chemical
  Society} \textbf{\bibinfo{volume}{138}}, \bibinfo{pages}{1065}
  (\bibinfo{year}{2016}).

\bibitem[{\citenamefont{Yeh and Chai}(2016)}]{yeh2016role}
\bibinfo{author}{\bibfnamefont{C.-N.} \bibnamefont{Yeh}} \bibnamefont{and}
  \bibinfo{author}{\bibfnamefont{J.-D.} \bibnamefont{Chai}}, ``Role of
  kekul{\'e} and non-kekul{\'e} structures in the radical character of
  alternant polycyclic aromatic hydrocarbons: a tao-dft study,''
  \bibinfo{journal}{Scientific reports} \textbf{\bibinfo{volume}{6}},
  \bibinfo{pages}{1} (\bibinfo{year}{2016}).

\bibitem[{\citenamefont{Kertesz}(1995)}]{kertesz1995structure}
\bibinfo{author}{\bibfnamefont{M.}~\bibnamefont{Kertesz}}, ``Structure and
  electronic structure of low-band-gap ladder polymers,''
  \bibinfo{journal}{Macromolecules} \textbf{\bibinfo{volume}{28}},
  \bibinfo{pages}{1475} (\bibinfo{year}{1995}).

\bibitem[{\citenamefont{Giannozzi et~al.}(2009)\citenamefont{Giannozzi, Baroni,
  Bonini, Calandra, Car, Cavazzoni, Ceresoli, Chiarotti, Cococcioni, Dabo
  et~al.}}]{giannozzi2009quantum}
\bibinfo{author}{\bibfnamefont{P.}~\bibnamefont{Giannozzi}},
  \bibinfo{author}{\bibfnamefont{S.}~\bibnamefont{Baroni}},
  \bibinfo{author}{\bibfnamefont{N.}~\bibnamefont{Bonini}},
  \bibinfo{author}{\bibfnamefont{M.}~\bibnamefont{Calandra}},
  \bibinfo{author}{\bibfnamefont{R.}~\bibnamefont{Car}},
  \bibinfo{author}{\bibfnamefont{C.}~\bibnamefont{Cavazzoni}},
  \bibinfo{author}{\bibfnamefont{D.}~\bibnamefont{Ceresoli}},
  \bibinfo{author}{\bibfnamefont{G.~L.} \bibnamefont{Chiarotti}},
  \bibinfo{author}{\bibfnamefont{M.}~\bibnamefont{Cococcioni}},
  \bibinfo{author}{\bibfnamefont{I.}~\bibnamefont{Dabo}}, \bibnamefont{et~al.},
  ``Quantum espresso: a modular and open-source software project for quantum
  simulations of materials,'' \bibinfo{journal}{Journal of Physics: Condensed
  Matter} \textbf{\bibinfo{volume}{21}}, \bibinfo{pages}{395502}
  (\bibinfo{year}{2009}).

\bibitem[{\citenamefont{Giannozzi et~al.}(2017)\citenamefont{Giannozzi,
  Andreussi, Brumme, Bunau, Nardelli, Calandra, Car, Cavazzoni, Ceresoli,
  Cococcioni et~al.}}]{giannozzi2017advanced}
\bibinfo{author}{\bibfnamefont{P.}~\bibnamefont{Giannozzi}},
  \bibinfo{author}{\bibfnamefont{O.}~\bibnamefont{Andreussi}},
  \bibinfo{author}{\bibfnamefont{T.}~\bibnamefont{Brumme}},
  \bibinfo{author}{\bibfnamefont{O.}~\bibnamefont{Bunau}},
  \bibinfo{author}{\bibfnamefont{M.~B.} \bibnamefont{Nardelli}},
  \bibinfo{author}{\bibfnamefont{M.}~\bibnamefont{Calandra}},
  \bibinfo{author}{\bibfnamefont{R.}~\bibnamefont{Car}},
  \bibinfo{author}{\bibfnamefont{C.}~\bibnamefont{Cavazzoni}},
  \bibinfo{author}{\bibfnamefont{D.}~\bibnamefont{Ceresoli}},
  \bibinfo{author}{\bibfnamefont{M.}~\bibnamefont{Cococcioni}},
  \bibnamefont{et~al.}, ``Advanced capabilities for materials modelling with
  quantum espresso,'' \bibinfo{journal}{Journal of Physics: Condensed Matter}
  \textbf{\bibinfo{volume}{29}}, \bibinfo{pages}{465901}
  (\bibinfo{year}{2017}).

\bibitem[{\citenamefont{Giannozzi et~al.}(2020)\citenamefont{Giannozzi,
  Baseggio, Bonfà, Brunato, Car, Carnimeo, Cavazzoni, de~Gironcoli, Delugas,
  Ferrari~Ruffino et~al.}}]{giannozzi2020}
\bibinfo{author}{\bibfnamefont{P.}~\bibnamefont{Giannozzi}},
  \bibinfo{author}{\bibfnamefont{O.}~\bibnamefont{Baseggio}},
  \bibinfo{author}{\bibfnamefont{P.}~\bibnamefont{Bonfà}},
  \bibinfo{author}{\bibfnamefont{D.}~\bibnamefont{Brunato}},
  \bibinfo{author}{\bibfnamefont{R.}~\bibnamefont{Car}},
  \bibinfo{author}{\bibfnamefont{I.}~\bibnamefont{Carnimeo}},
  \bibinfo{author}{\bibfnamefont{C.}~\bibnamefont{Cavazzoni}},
  \bibinfo{author}{\bibfnamefont{S.}~\bibnamefont{de~Gironcoli}},
  \bibinfo{author}{\bibfnamefont{P.}~\bibnamefont{Delugas}},
  \bibinfo{author}{\bibfnamefont{F.}~\bibnamefont{Ferrari~Ruffino}},
  \bibnamefont{et~al.}, ``Quantum espresso toward the exascale,''
  \bibinfo{journal}{The Journal of Chemical Physics}
  \textbf{\bibinfo{volume}{152}}, \bibinfo{pages}{154105}
  (\bibinfo{year}{2020}).

\bibitem[{\citenamefont{Neese}(2012)}]{orcageneric}
\bibinfo{author}{\bibfnamefont{F.}~\bibnamefont{Neese}}, ``The orca program
  system,'' \bibinfo{journal}{WIREs Computational Molecular Science}
  \textbf{\bibinfo{volume}{2}}, \bibinfo{pages}{73} (\bibinfo{year}{2012}).

\bibitem[{\citenamefont{Perdew et~al.}(1996{\natexlab{a}})\citenamefont{Perdew,
  Burke, and Ernzerhof}}]{perdew1996generalized}
\bibinfo{author}{\bibfnamefont{J.~P.} \bibnamefont{Perdew}},
  \bibinfo{author}{\bibfnamefont{K.}~\bibnamefont{Burke}}, \bibnamefont{and}
  \bibinfo{author}{\bibfnamefont{M.}~\bibnamefont{Ernzerhof}}, ``Generalized
  gradient approximation made simple,'' \bibinfo{journal}{Physical Review
  Letters} \textbf{\bibinfo{volume}{77}}, \bibinfo{pages}{3865}
  (\bibinfo{year}{1996}{\natexlab{a}}).

\bibitem[{\citenamefont{Perdew et~al.}(1996{\natexlab{b}})\citenamefont{Perdew,
  Ernzerhof, and Burke}}]{hybrids1}
\bibinfo{author}{\bibfnamefont{J.~P.} \bibnamefont{Perdew}},
  \bibinfo{author}{\bibfnamefont{M.}~\bibnamefont{Ernzerhof}},
  \bibnamefont{and} \bibinfo{author}{\bibfnamefont{K.}~\bibnamefont{Burke}},
  ``Rationale for mixing exact exchange with density functional
  approximations,'' \bibinfo{journal}{The Journal of Chemical Physics}
  \textbf{\bibinfo{volume}{105}}, \bibinfo{pages}{9982}
  (\bibinfo{year}{1996}{\natexlab{b}}).

\bibitem[{\citenamefont{Adamo and Barone}(1999)}]{PBE0hybrid}
\bibinfo{author}{\bibfnamefont{C.}~\bibnamefont{Adamo}} \bibnamefont{and}
  \bibinfo{author}{\bibfnamefont{V.}~\bibnamefont{Barone}}, ``Toward reliable
  density functional methods without adjustable parameters: The pbe0 model,''
  \bibinfo{journal}{The Journal of Chemical Physics}
  \textbf{\bibinfo{volume}{110}}, \bibinfo{pages}{6158} (\bibinfo{year}{1999}).

\bibitem[{\citenamefont{Yazyev}(2010)}]{Yazyev_2010}
\bibinfo{author}{\bibfnamefont{O.~V.} \bibnamefont{Yazyev}}, ``Emergence of
  magnetism in graphene materials and nanostructures,''
  \bibinfo{journal}{Reports on Progress in Physics}
  \textbf{\bibinfo{volume}{73}}, \bibinfo{pages}{056501}
  (\bibinfo{year}{2010}).

\bibitem[{\citenamefont{Majumdar and Ghosh}(1969)}]{majumdar1969next}
\bibinfo{author}{\bibfnamefont{C.~K.} \bibnamefont{Majumdar}} \bibnamefont{and}
  \bibinfo{author}{\bibfnamefont{D.~K.} \bibnamefont{Ghosh}}, ``On
  next-nearest-neighbor interaction in linear chain. i,''
  \bibinfo{journal}{Journal of Mathematical Physics}
  \textbf{\bibinfo{volume}{10}}, \bibinfo{pages}{1388} (\bibinfo{year}{1969}).

\bibitem[{\citenamefont{Majumdar}(1970)}]{majumdar1970antiferromagnetic}
\bibinfo{author}{\bibfnamefont{C.~K.} \bibnamefont{Majumdar}},
  ``Antiferromagnetic model with known ground state,''
  \bibinfo{journal}{Journal of Physics C: Solid State Physics}
  \textbf{\bibinfo{volume}{3}}, \bibinfo{pages}{911} (\bibinfo{year}{1970}).

\bibitem[{\citenamefont{Haldane}(1982)}]{HaldaneMG}
\bibinfo{author}{\bibfnamefont{F.~D.~M.} \bibnamefont{Haldane}}, ``Spontaneous
  dimerization in the $s=1/2$ heisenberg antiferromagnetic chain with competing
  interactions,'' \bibinfo{journal}{Phys. Rev. B}
  \textbf{\bibinfo{volume}{25}}, \bibinfo{pages}{4925} (\bibinfo{year}{1982}).

\bibitem[{\citenamefont{Van~den Broek}(1980)}]{van1980exact}
\bibinfo{author}{\bibfnamefont{P.}~\bibnamefont{Van~den Broek}}, ``Exact value
  of the ground state energy of the linear antiferromagnetic heisenberg chain
  with nearest and next-nearest neighbour interactions,''
  \bibinfo{journal}{Physics letters A} \textbf{\bibinfo{volume}{77}},
  \bibinfo{pages}{261} (\bibinfo{year}{1980}).

\bibitem[{\citenamefont{Xu et~al.}(2021)\citenamefont{Xu, Zhang, Guo, and
  Gong}}]{spindimers}
\bibinfo{author}{\bibfnamefont{H.-Z.} \bibnamefont{Xu}},
  \bibinfo{author}{\bibfnamefont{S.-Y.} \bibnamefont{Zhang}},
  \bibinfo{author}{\bibfnamefont{G.-C.} \bibnamefont{Guo}}, \bibnamefont{and}
  \bibinfo{author}{\bibfnamefont{M.}~\bibnamefont{Gong}}, ``Exact dimer phase
  with anisotropic interaction for one dimensional magnets,''
  \bibinfo{journal}{Scientific Reports} \textbf{\bibinfo{volume}{11}}
  (\bibinfo{year}{2021}).

\bibitem[{\citenamefont{Furukawa et~al.}(2012)\citenamefont{Furukawa, Sato,
  Onoda, and Furusaki}}]{furusakichain}
\bibinfo{author}{\bibfnamefont{S.}~\bibnamefont{Furukawa}},
  \bibinfo{author}{\bibfnamefont{M.}~\bibnamefont{Sato}},
  \bibinfo{author}{\bibfnamefont{S.}~\bibnamefont{Onoda}}, \bibnamefont{and}
  \bibinfo{author}{\bibfnamefont{A.}~\bibnamefont{Furusaki}}, ``Ground-state
  phase diagram of a s=1/2 frustrated ferromagnetic xxz chain: Haldane dimer
  phase and gapped/gapless chiral phases,'' \bibinfo{journal}{Phys. Rev. B}
  \textbf{\bibinfo{volume}{86}}, \bibinfo{pages}{094417}
  (\bibinfo{year}{2012}).

\bibitem[{\citenamefont{Agrapidis et~al.}(2019)\citenamefont{Agrapidis,
  Drechsler, van~den Brink, and Nishimoto}}]{agrapidis2019coexistence}
\bibinfo{author}{\bibfnamefont{C.~E.} \bibnamefont{Agrapidis}},
  \bibinfo{author}{\bibfnamefont{S.-L.} \bibnamefont{Drechsler}},
  \bibinfo{author}{\bibfnamefont{J.}~\bibnamefont{van~den Brink}},
  \bibnamefont{and}
  \bibinfo{author}{\bibfnamefont{S.}~\bibnamefont{Nishimoto}}, ``Coexistence of
  valence-bond formation and topological order in the frustrated ferromagnetic
  $ j\_1 $-$ j\_2 $ chain,'' \bibinfo{journal}{SciPost Physics}
  \textbf{\bibinfo{volume}{6}}, \bibinfo{pages}{019} (\bibinfo{year}{2019}).

\bibitem[{\citenamefont{Affleck et~al.}(1988)\citenamefont{Affleck, Kennedy,
  Lieb, and Tasaki}}]{akltpaper}
\bibinfo{author}{\bibfnamefont{I.}~\bibnamefont{Affleck}},
  \bibinfo{author}{\bibfnamefont{T.}~\bibnamefont{Kennedy}},
  \bibinfo{author}{\bibfnamefont{E.~H.} \bibnamefont{Lieb}}, \bibnamefont{and}
  \bibinfo{author}{\bibfnamefont{H.}~\bibnamefont{Tasaki}}, ``Valence bond
  ground states in isotropic quantum antiferromagnets,''
  \bibinfo{journal}{Communications in Mathematical Physics}
  \textbf{\bibinfo{volume}{115}}, \bibinfo{pages}{477} (\bibinfo{year}{1988}).

\bibitem[{\citenamefont{Villain et~al.}(1980)\citenamefont{Villain, Bidaux,
  Carton, and Conte}}]{villain1980order}
\bibinfo{author}{\bibfnamefont{J.}~\bibnamefont{Villain}},
  \bibinfo{author}{\bibfnamefont{R.}~\bibnamefont{Bidaux}},
  \bibinfo{author}{\bibfnamefont{J.-P.} \bibnamefont{Carton}},
  \bibnamefont{and} \bibinfo{author}{\bibfnamefont{R.}~\bibnamefont{Conte}},
  ``Order as an effect of disorder,'' \bibinfo{journal}{Journal de Physique}
  \textbf{\bibinfo{volume}{41}}, \bibinfo{pages}{1263} (\bibinfo{year}{1980}).

\bibitem[{\citenamefont{Shimizu et~al.}(2014)\citenamefont{Shimizu, Nobusue,
  Miyoshi, and Tobe}}]{tobepure}
\bibinfo{author}{\bibfnamefont{A.}~\bibnamefont{Shimizu}},
  \bibinfo{author}{\bibfnamefont{S.}~\bibnamefont{Nobusue}},
  \bibinfo{author}{\bibfnamefont{H.}~\bibnamefont{Miyoshi}}, \bibnamefont{and}
  \bibinfo{author}{\bibfnamefont{Y.}~\bibnamefont{Tobe}}, ``Indenofluorene
  cogeners: Biradicaloids and beyond,'' \bibinfo{journal}{Pure Appl. Chem.}
  \textbf{\bibinfo{volume}{86}} (\bibinfo{year}{2014}).

\bibitem[{\citenamefont{Dressler et~al.}(2017)\citenamefont{Dressler, Zhou,
  Marshall, Kishi, Takamuku, Wei, Spisak, Nakano, Petrukhina, and
  Haley}}]{dressler2017synthesis}
\bibinfo{author}{\bibfnamefont{J.~J.} \bibnamefont{Dressler}},
  \bibinfo{author}{\bibfnamefont{Z.}~\bibnamefont{Zhou}},
  \bibinfo{author}{\bibfnamefont{J.~L.} \bibnamefont{Marshall}},
  \bibinfo{author}{\bibfnamefont{R.}~\bibnamefont{Kishi}},
  \bibinfo{author}{\bibfnamefont{S.}~\bibnamefont{Takamuku}},
  \bibinfo{author}{\bibfnamefont{Z.}~\bibnamefont{Wei}},
  \bibinfo{author}{\bibfnamefont{S.~N.} \bibnamefont{Spisak}},
  \bibinfo{author}{\bibfnamefont{M.}~\bibnamefont{Nakano}},
  \bibinfo{author}{\bibfnamefont{M.~A.} \bibnamefont{Petrukhina}},
  \bibnamefont{and} \bibinfo{author}{\bibfnamefont{M.~M.} \bibnamefont{Haley}},
  ``Synthesis of the unknown indeno [1, 2-a] fluorene regioisomer:
  Crystallographic characterization of its dianion,''
  \bibinfo{journal}{Angewandte Chemie} \textbf{\bibinfo{volume}{129}},
  \bibinfo{pages}{15565} (\bibinfo{year}{2017}).

\bibitem[{\citenamefont{Shimizu et~al.}(2013)\citenamefont{Shimizu, Kishi,
  Nakano, Shiomi, Sato, Takui, Hisaki, Miyata, and Tobe}}]{shimizu2013indeno}
\bibinfo{author}{\bibfnamefont{A.}~\bibnamefont{Shimizu}},
  \bibinfo{author}{\bibfnamefont{R.}~\bibnamefont{Kishi}},
  \bibinfo{author}{\bibfnamefont{M.}~\bibnamefont{Nakano}},
  \bibinfo{author}{\bibfnamefont{D.}~\bibnamefont{Shiomi}},
  \bibinfo{author}{\bibfnamefont{K.}~\bibnamefont{Sato}},
  \bibinfo{author}{\bibfnamefont{T.}~\bibnamefont{Takui}},
  \bibinfo{author}{\bibfnamefont{I.}~\bibnamefont{Hisaki}},
  \bibinfo{author}{\bibfnamefont{M.}~\bibnamefont{Miyata}}, \bibnamefont{and}
  \bibinfo{author}{\bibfnamefont{Y.}~\bibnamefont{Tobe}}, ``Indeno [2, 1-b]
  fluorene: A 20-$\pi$-electron hydrocarbon with very low-energy light
  absorption,'' \bibinfo{journal}{Angewandte Chemie}
  \textbf{\bibinfo{volume}{125}}, \bibinfo{pages}{6192} (\bibinfo{year}{2013}).

\bibitem[{\citenamefont{White and Affleck}(1996)}]{white1996dimerization}
\bibinfo{author}{\bibfnamefont{S.~R.} \bibnamefont{White}} \bibnamefont{and}
  \bibinfo{author}{\bibfnamefont{I.}~\bibnamefont{Affleck}}, ``Dimerization and
  incommensurate spiral spin correlations in the zigzag spin chain: Analogies
  to the kondo lattice,'' \bibinfo{journal}{Physical Review B}
  \textbf{\bibinfo{volume}{54}}, \bibinfo{pages}{9862} (\bibinfo{year}{1996}).

\end{thebibliography}

\end{document}